\documentclass[]{spie}  

\usepackage{amsmath,amsfonts,amssymb}
\usepackage{graphicx}
\usepackage[colorlinks=true, allcolors=blue]{hyperref}

\usepackage{tcolorbox,bm}
\DeclareMathOperator*{\argmin}{\arg\!\min}

\usepackage{color}

\newcommand\figtwo{5.5cm}
\newcommand\figthree{4.2cm}
\newcommand\figthreealt{4.1cm}

\graphicspath{{figs/}}

\title{Iterative and greedy algorithms for the sparsity in levels model in compressed sensing}

\author[a]{Ben Adcock}
\author[a]{Simone Brugiapaglia}
\author[a]{Matthew King-Roskamp}
\affil[a]{Department of Mathematics, Simon Fraser University, Burnaby, B.C., Canada. }

\authorinfo{E-mail: ben\_adcock@sfu.ca, simone\_brugiapaglia@sfu.ca, mkingros@sfu.ca \\ Send correspondence to M.\ King--Roskamp}

\begin{document}

\newpage

\maketitle

\begin{abstract}
Motivated by the question of optimal functional approximation via compressed sensing, we propose generalizations of the Iterative Hard Thresholding and the Compressive Sampling Matching Pursuit algorithms able to  promote sparse in levels signals. We show, by means of numerical experiments, that the proposed algorithms are successfully able to outperform their unstructured variants when the signal exhibits the sparsity structure of interest. Moreover, in the context of piecewise smooth function approximation, we numerically demonstrate that the structure promoting decoders outperform their unstructured variants and the basis pursuit program when the encoder is structure agnostic.
\end{abstract}


\section{Introduction}

In classical compressed sensing, one considers the recovery of an $s$-sparse vector  $x \in \mathbb{C}^N$ from noisy measurements $y = A x + e \in \mathbb{C}^m$.  It is now well understood that classical sparsity -- where the vector $x$ has at most $s$ nonzero components -- is but one low-dimensional signal model, and that more sophisticated models may bring significant performance gains in practice \cite{DuarteEldarStructuredCS,BaranuikModelCS,TraonmilinGribonvalRIP}.  Well known \textit{structured sparsity} models include group or block sparsity, joint sparsity, weighted sparsity, connected tree sparsity and numerous others.

The focus of this paper is the so-called \textit{local sparsity in levels} model\cite{AHPRBreaking}.  In this model, a vector is divided in $r$ disjoint \textit{levels} and separate sparsity allowed within each.  Thus, one now has a vector $\bm{s} = (s_1,\ldots,s_r)$ of sparsities as opposed to a single sparsity $s$.  While simple, this model plays a crucial role in compressed sensing for imaging, where the local sparsities are typically related to the wavelet scales, and the target vector $x$ represents the approximately sparse wavelet coefficients of an image.  The sparsity in levels model, and its corresponding compressed sensing theory, allows one to design sampling strategies which leverage the characteristic local sparsity structure of images (so-called \textit{asymptotic sparsity}) thus giving significant practical benefits gains\cite{AsymptoticCS,AHPRBreaking,OptimalSamplingQuest}.  

Imaging aside, the sparsity in levels model also arises naturally in other contexts.  For instance, it can be used to model so-called \textit{sparse and distributed} or \textit{sparse and balanced} vectors, which occur in parallel acquisition problems\cite{AdcockChunParallel,Chun&Adcock:16ITW} and radar\cite{Dorsch2016}.  The specific case of two levels also arises in the \textit{sparse corruptions} problem \cite{BAEtAlCorruptions,LiCorruptionsConstrApprox}.

The focus of past work on this model has been on convex optimization-based decoders such as \textit{Quadratically-Constrained Basis Pursuit (QCBP)}
\begin{equation}
\label{QCBP}
\min_{z \in \mathbb{C}^N} \Vert z \Vert_{\ell^1}\ \mbox{subject to $\Vert A z  -y \Vert_{\ell^2} \leq \eta$},
\end{equation}
or closely-related weighted variants\cite{TraonmilinGribonvalRIP}, with the weights being used to promote the sparsity in levels structure.  Both uniform\cite{LiAdcockRIP} and nonuniform\cite{AHPRBreaking} recovery guarantees have been established for this decoder, with measurement conditions relating the number of measurements $m$ to the local sparsities $\bm{s}$.  However, and perhaps surprisingly, no attention has been paid to alternatives to convex optimization -- namely, greedy and iterative methods -- despite these being quite widely used and studied both in the case of classical sparsity\cite{FoucartRauhutCSbook} and various other structured structured sparsity models\cite{BaranuikModelCS}.

In this paper, we introduce and study generalizations of the classical \textit{Iterative Hard Thresholding (IHT)} and \textit{Compressive Sampling Matching Pursuit (CoSaMP)} algorithms for the sparsity in levels model, known as \textit{IHT in Levels (IHTL)} and \textit{CoSaMP in Levels (CoSaMPL)} respectively.  These generalizations are natural, and straightforward to implement.  We then present a series of numerical experiments demonstrating the benefits of these decoders in the presence of sparsity in levels.  Specifically, we highlight several settings in which promoting this additional structure leads to better recovery over the classical IHT and CoSaMP algorithms.  The purpose of this paper is to establish this proof of concept.  We defer a full theoretical analysis to an upcoming work.  However, to provide some context, we do briefly discuss the relevant theoretical tools needed to analyze sparsity in levels and highlight existing results for the QCBP decoder.

This work was motivated by the question of optimal function approximation via compressed sensing\cite{BASBMKRCSwavelet}.  Specifically, given a function class $F$ for which it is known the best $s$-term approximation rate decays like $s^{-\alpha}$ for some $\alpha > 0$, can one design $m$ compressed sensing measurements and a decoder so that the resulting approximation achieves an $\mathcal{O}(m^{-\alpha})$ error?  Moreover, can this be achieved by a \textit{black box}, with a decoder that has polynomial runtime in $m$?  The first question has been answered affirmatively for the class of piecewise $\alpha$-H\"older functions of one variable\cite{BASBMKRCSwavelet}.  Note that sparsity in level is crucial for obtaining the optimal error rate.  However, the decoder is based on a weighted $\ell^1$ minimization program\cite{BASBMKRCSwavelet}, and thus does not give an affirmative answer to the second question.  Striving for a true black box motivates one to consider non-optimization based approaches, such as the IHTL and CoSaMPL algorithms we introduce in this paper. With this motivation in mind, we conclude the paper with some experiments on approximation of piecewise smooth functions via compressed sensing, comparing the introduced algorithms to \eqref{QCBP}.



\section{Classical compressed sensing}

Recall that a vector $x = (x_i)^{N}_{i=1} \in \mathbb{C}^N$ is \textit{$s$-sparse} if it has at most $1 \leq s \leq N$ nonzero entries: that is,
\begin{equation*}
| \mathrm{supp}(x) | \leq s,    
\end{equation*}
where $\mathrm{supp}(x) = \{ i : x_i \neq 0 \}$ is the \textit{support} of $x$.  Classical compressed sensing concerns the recovery of a sparse vector $x$ from $m$ noisy linear measurements
\begin{equation*}
y = A x + e \in \mathbb{C}^m,    
\end{equation*}
where $A \in \mathbb{C}^{m\times N}$ is the measurement matrix and $e \in \mathbb{C}^m$ is a noise vector.

\subsection{IHT and CoSaMP}

For a vector $x \in \mathbb{C}^{N}$ (not necessarily sparse), let $L_{s}(x)$ be the index set of its $s$ largest entries in absolute value.  The \textit{hard thresholding} operator $H_{s} : \mathbb{bbC}^N \rightarrow \mathbb{C}^N$ is, for $x = (x_i)^{N}_{i=1} \in \mathbb{C}^{N}$,  defined by
\begin{equation*}
H_{s}(x) = (H_{s}(x)_{i})_{i=1}^{N},\qquad    H_{s}(x)_{i} = \begin{cases} x_{i} &i \in L_{s}(x) \\
    0 &\text{otherwise}
        \end{cases}.
\end{equation*}
That is, $H_{s}(x)$ is the vector of the $s$ largest entries of $x$ with all other entries set to zero.  The classical \textit{Iterative Hard Thresholding (IHT)} algorithm is now defined as follows:

\begin{tcolorbox}
Function $\mathrm{IHT}(A,y,s)$ 
\\
\noindent \textbf{Inputs:} $A \in \mathbb{C}^{m\times N}$, $y \in \mathbb{C}^m$, sparsity $s$ 
\\
\noindent \textbf{Initialization:} $x^{(0)} \in \mathbb{C}^N$ (e.g.\ $x^{(0)} = 0$)
\\
\noindent \textbf{Iterate:} Until some stopping criterion is met at $n = \overline{n}$, set \\
\begin{equation*}
    x^{(n+1)}= H_{s}(x^{(n)} + A^{*}(y - Ax^{(n)}))
\end{equation*}

\noindent \textbf{Output:} $\hat{x} = x^{(\overline{n})}$
\end{tcolorbox}

The idea of the IHT algorithm is to combine the classical Landweber iteration\cite{Landweber1951} with the action of a hard thresholding operator to promote sparse solutions.  Specifically, IHT corresponds to a Landweber iteration with constant unit step size followed by a pruning operation performed via hard thresholding. The power of IHT lies within the extreme simplicity and efficiency of its iteration.

The ancestors of IHT are the \emph{Iterated Shrkinage} methods\cite{elad2007wide} and IHT was introduced in the context of compressed sensing in the late 2000s\cite{BlumensathDavies2008,BlumensathDavies2009}. In the first version of the IHT algorithm\cite{BlumensathDavies2008,BlumensathDavies2009} the  step size of the Landweber iteration is constant with respect to the iteration. Accelerated versions of IHT with variable step size were introduced later.\cite{BlumensathDavies2010, Blumensath2012} A generalization of IHT to the union of subspaces model was studied in the context of model-based compressed sensing.\cite{BaranuikModelCS,HedgeIndykSchmidt2015}

The \textit{Compressive Sampling Matching Pursuity (CoSaMP)} algorithm is:

\begin{tcolorbox}
Function $\mathrm{CoSaMP}(A,y,s)$ \\
\noindent \textbf{Inputs:} $A \in \mathbb{C}^{m\times N}$, $y \in \mathbb{C}^m$, sparsity $s$ \\
\noindent \textbf{Initialization:} $x^{(0)} \in \mathbb{C}^N$ (e.g.\ $x^{(0)} = 0$)
\\
\noindent \textbf{Iterate:} Until some stopping criterion is met at $n = \overline{n}$, set 
\begin{align*}
    U^{(n+1)} &= \text{supp}(x^{(n+1)}) \cup L_{2s}(A^{*}(y-Ax^{(n)})) \\
    u^{(n+1)} &\in \argmin_{z \in \mathbb{C}^N} \{ \Vert y -Az \Vert_{2} \: : \: \text{supp}(z) \subset U^{(n+1)} \} \\
    x^{(n+1)} &= H_{s}(u^{(n+1)})
\end{align*}

\noindent \textbf{Output:} $\hat{x} = x^{(\overline{n})}$
\end{tcolorbox}

CoSaMP was proposed in the late 2000's\cite{NeedellTropp2008}, inspired by the so-called \emph{regularized orthogonal matching pursuit} algorithm.\cite{NeedellVershynin2009,NeedellVershynin2010} CoSaMP combines the principles of greedy (multiple) index selection and of orthogonal projection used in orthogonal matching pursuit with hard thresholding.  It takes advantage of a greedy index selection principle typical of matching pursuit algorithms by looking for the $2s$ columns of $A$ that are most correlated with the residual. After updating the support accordingly, CoSaMP performs a least-squares  projection onto the active support followed by a hard thresholding to preserve sparsity. Similary to IHT, extensions of CoSaMP to the union of subspaces model were developed and analyzed in the context of model-based compressed sensing.\cite{BaranuikModelCS, HedgeIndykSchmidt2015}

A potentially attractive feature of both the IHT and CoSaMP algorithms is that the reconstruction $\hat{x}$ is exactly $s$-sparse.  This is not the case in general when $\hat{x}$ is a minimizer of \eqref{QCBP}, i.e.
\begin{equation*}
\hat{x} \in \argmin_{z \in \mathbb{C}^N} \Vert z \Vert_{\ell^1}\ \mbox{subject to $\Vert A z  -y \Vert_{\ell^2} \leq \eta$}.
\end{equation*}
Note that IHT and CoSaMP require knowledge of $s$, but no knowledge of the noise level $\Vert e \Vert_{\ell^2}$, whereas \eqref{QCBP} requires no knowledge of $s$ but knowledge of $\Vert e \Vert_{\ell^2}$. In fact, although stable and robust recovery guarantees for QCBP can be shown for $\eta \geq \|e\|_{\ell^2} $ (under the restricted isometry property) and for $0 \leq \eta < \|e\|_{\ell^2}$ (under the restricted isometry and the quotient properties), the optimal parameter tuning strategy for QCBP is $\eta = \|e\|_{\ell^2}$.\cite{BrugiapagliaAdcock2018, Wojtaszczyk2010}


\subsection{Recovery guarantees}



Much as with \eqref{QCBP}, recovery guarantees for IHT and CoSaMP are typically based on the RIP:

\begin{definition}
Let $1 \leq s \leq N$.  The $s^{\text{th}}$ \textit{Restricted Isometry Constant (RIC)} $\delta_s$ of a matrix $A \in \mathbb{C}^{m \times N}$ is the smallest $\delta \geq 0$ such that
\begin{equation}
\label{RIP}
(1-\delta) \| x \|^2_{\ell^2} \leq \| A x \|^2_{\ell^2} \leq (1+\delta) \| x \|^2_{\ell^2},\quad \mbox{for all $s$-sparse $x$}.
\end{equation}
If $0 < \delta_s < 1$ then $A$ is said to have the \textit{Restricted Isometry Property (RIP)} of order $s$.
\end{definition}




\begin{theorem}
Suppose that the $6s$-th RIC constant of $A \in \mathbb{C}^{m \times N}$ satisfies $\delta_{6s} < \frac{1}{\sqrt{3}}$. Then, for all $x \in \mathbb{C}^{N}, e \in \mathbb{C}^{m}$, the sequence $x^{(n)}$ defined by $\mathrm{IHT}(A,y,2s)$ with $y = A x + e$ and $x^{(0)} = 0$ satisfies, for any $n \geq 0$,
\begin{align*}
 \Vert x - x^{(n)} \Vert_{\ell^{1}} &\leq C\sigma_{s}(x)_{\ell^{1}} + D \sqrt{s} \Vert e \Vert_{2} + 2 \sqrt{s} \rho^{n} \Vert x \Vert_{2}, \\
\Vert x - x^{(n)} \Vert_{\ell^{2}} &\leq \frac{C}{\sqrt{s}}\sigma_{s}(x)_{\ell^{1}} + D \Vert e \Vert_{2} + \rho^{n} \Vert x \Vert_{2},
\end{align*}
where $\rho = \sqrt{3}\delta_{6s} < 1$, and $C,D >0$ are constants only depending on $\delta_{6s}$.
\end{theorem}

\begin{theorem}  Suppose that the $8s$-th RIC constant of $A$ satisfies
\begin{equation*}
    \delta_{8s} < \frac{\sqrt{\frac{11}{3}} - 1}{4} \approx 0.478.
\end{equation*}
Then for $x \in \mathbb{C}^{N}$, $e \in \mathbb{C}^{m}$ the sequence $x^{(n)}$ defined by $\mathrm{CoSaMP}(A,y,2s)$ with $y=Ax + e$ and $x^{(0)} = 0$, satisfies for any $n \geq 0$,
\begin{align*}
    \Vert x - x^{(n)} \Vert_{\ell^{1}} &\leq C\sigma_{s}(x)_{\ell^1} + D \sqrt{s} \Vert e \Vert_{2} + 2 \sqrt{s} \rho^{n} \Vert x \Vert_{2}, \\
\Vert x - x^{(n)} \Vert_{2} &\leq \frac{C}{\sqrt{s}}\sigma_{s}(x)_{\ell^{1}} + D\Vert e \Vert_{2} + 2 \rho^{n} \Vert x \Vert_{2},
\end{align*}
where  $\rho = \sqrt{\frac{2\delta_{8s}^{2}(1+3\delta_{8s}^{2})}{1-\delta_{4s}^{2}}} < 1$ and $C,D >0$ are constants only depending on $\delta_{8s}$.
\end{theorem}

See \cite[Thms.\ 6.21 \& 6.28]{FoucartRauhutCSbook} respectively.  Here $\sigma_{s}(x)_{\ell^1}$ is the $\ell^1$-norm best $s$-term approximation error:
\begin{equation*}
 \sigma_{s}(x)_{\ell^1} = \min \left \{ \Vert x - z \Vert_{\ell^1}  : z \text{ is $s$-sparse} \right \}.  
\end{equation*}

\section{Compressed sensing with local structure}

We now consider the local sparsity in levels model.

\subsection{Local sparsity in levels}

The sparsity in levels model divides a vector $x$ into $r$ separate levels, and then separately measures the sparsity within each one:

\begin{definition}
 Let $r \geq 1$, $\bm{M} = (M_1,\ldots,M_r)$, where $1 \leq M_1 < M_2 < \ldots < M_r = N$ and $\bm{s} = (s_1,\ldots,s_r)$, where $s_k \leq M_k - M_{k-1}$ for $k=1,\ldots,r$, with $M_0 = 0$. A vector $x = (x_i)^{N}_{i=1} \in \mathbb{C}^N$ is \textit{$(\bm{s},\bm{M})$-sparse} if
\begin{equation}
\left | \text{supp}(x) \cap \{ M_{k-1}+1,\ldots,M_k \} \right | \leq s_k,\quad k = 1,\ldots,r.
\end{equation}
We write $\Sigma_{\bm{s},\bm{M}} \subseteq \mathbb{C}^M$ for the set of $(\bm{s},\bm{M})$-sparse vectors. 
We refer to $\bm{M} = (M_1,\ldots,M_r)$ as \textit{sparsity levels}, and $\bm{s} = (s_1,\ldots,s_r)$ as \textit{local sparsities} respectively.
\end{definition}

In imaging applications, the levels typically correspond to wavelet scales, in which case the $k^{\text{th}}$ level has size roughly $2^k$.  In other applications, for instance, parallel acquisition problems, the levels are typically equally sized.  In function approximation, as we will consider later, it is typical to consider an $s$-sparse vector with a two level structure based on $s_1 = M_1 = s/2$ and $s_2 = s/2$.  That is, the first $s/2$ coefficients are nonzero, and the remaining $s/2$ can be arbitrarily location within the indices $\{s/2+1,\ldots,N\}$.

\subsection{Structured sampling or structured recovery}

Having defined this structured sparsity model, we need a mechanism to exploit it.  These fall into two broad categories.  In \textit{structured sampling} one seeks to design measurements to promote the given structure.  Conversely, in \textit{structured recovery} one designs a decoder to promote the structure.  Note that the former is easily achieved, at least in theory.  One simply constructs $A \in \mathbb{C}^{m \times N}$ as the block diagonal matrix whose $k^{\text{th}}$ block $A_k$, corresponding to the $k^{\text{th}}$ sparsity level, is an $m_k \times (M_k - M_{k-1})$ standard compressed sensing matrix.  Recovery can then be achieved in a level-by-level manner using standard decoders.  Of course, such a construction is generally not possible in practice, when the measurements are constrained by the physical sensing device (e.g.\ Fourier measurements in imaging applications).

\subsection{The Restricted Isometry Property in Levels}
\label{sec:RIPL}

In the sparsity in levels setting, the standard tool for establishing uniform recovery guarantees is the Restricted Isometry Property in Levels\cite{BastounisHansen}:

\begin{definition}
Let $\bm{M} = (M_1,\ldots,M_r)$ be sparsity levels and $\bm{s} = (s_1,\ldots,s_r)$ be local sparsities.  The $(\bm{s},\bm{M})^{\text{th}}$ \textit{Restricted Isometry Constant in Levels (RICL)} $\delta_{\bm{s},\bm{M}}$ of a matrix $A \in \mathbb{C}^{m \times N}$ is the smallest $\delta \geq 0$ such that
\begin{equation}
(1-\delta) \Vert x \Vert^2_{\ell^2} \leq \Vert A x \Vert^2_{\ell^2} \leq (1+\delta) \Vert  x \Vert^2_{\ell^2},\quad \forall x \in \Sigma_{\bm{s},\bm{M}}.
\end{equation}
If $0 < \delta_{\bm{s},\bm{M}} < 1$ then the matrix is said to have the \textit{Restricted Isometry Property in Levels (RIPL)} of order $(\bm{s},\bm{M})$.
\end{definition}

Analogously to the classical setting where the RIP is used to guarantee recovery, the RIPL is sufficient for recovery with appropriate decoders.  Specifically, if a matrix $A$ has the RIPL of suitable order, then stable and robust is ensured for the weighted QCBP decoder
\begin{equation*}
\min_{z \in \mathbb{C}^N} \Vert z \Vert_{\ell^1_w}\ \mbox{subject to $\Vert A z  -y \Vert_{\ell^2} \leq \eta$},   
\end{equation*}
where the weights $w = (w_i)^{N}_{i=1}$ are given by $w_i = \sqrt{s/s_k}$ for $M_{k-1} < i \leq M_k$\cite{TraonmilinGribonvalRIP}.  Note that this is a type of structured decoder.  One may also employ the unstructured unweighted QCBP decoder \eqref{QCBP}, provided the local sparsities do not differ too greatly, i.e.\ $\min \{s_k \} \approx \max \{s_k \}$\cite{BastounisHansen}.

Measurement matrices that satisfy the RIPL can be readily designed.
Note that a random matrix with independent normal entries having mean zero and variance $1/\sqrt{m}$ has the RIPL of order $\delta_{\bm{s},\bm{M}} \leq \delta$ with probability at least $1-\epsilon$, provided
\begin{equation*}
m \geq C \delta^{-2} \left ( \sum^{r}_{k=1} s_k \log \left ( \frac{\mathrm{e} (M_k - M_{k-1}) }{s_k} \right ) + \log(\epsilon^{-1}) \right ). 
\end{equation*}
Measurement conditions have also be shown for subsampled unitary matrices\cite{LiAdcockRIP}. Specifically, let $U \in \mathbb{C}^{N \times N}$ be unitary.  Let $\bm{N} = (N_1,\ldots,N_r)$ be a vector of \textit{sampling levels}, where $1 \leq N_1 < \cdots < N_r = N$, and $\bm{m} = (m_1,\ldots,m_r)$ be a vector of local numbers of measurements, where $m_k \leq N_k - N_{k-1}$ and $N_0 = 1$.  An \textit{$(\bm{m},\bm{N})$-multilevel random sampling} scheme is a set $\Omega = \Omega_1 \cup \cdots \cup \Omega_r$, where $\Omega \subseteq \{N_{k-1} + 1,\ldots,N_k \}$ is defined as follows: if $m_k = N_k - N_{k-1}$ then $\Omega_k = \{ N_{k-1}+1,\ldots,N_k \}$, otherwise $\Omega_k$ consists of $m_k$ values chosen uniformly and independently from $\{ N_{k-1}+1,\ldots,N_k \}$.  Consider the measurement matrix
\begin{equation*}
A = P_{\Omega} D U \in \mathbb{C}^{m \times N},
\end{equation*}
where $P_{\Omega}$ is the \textit{row selector} matrix, picking rows of $U$ corresponding to indices in $\Omega$, and $D$ is a diagonal scaling matrix with $i^{\text{th}}$ entry $\sqrt{\frac{N_k - N_{k-1}}{m_k}}$ if $N_{k-1}+1 < i \leq N_k$.  Then $A$ has the RIPL of order $\delta_{\bm{s},\bm{M}} \leq \delta$ with probability at least $1-\epsilon$, provided
\begin{equation*}
m_k \geq C \delta^{-2} (N_k - N_{k-1}) \left ( \sum^{r}_{k=1} \mu_{k,l} s_l \right ) \left ( r \log^2(s) \log(m) \log(N) + \log(\epsilon^{-1}) \right ),\quad k =1,\ldots,r.   
\end{equation*}
Here $m = m_1+\ldots+m_r$, $s = s_1+\ldots+s_r$ and $\mu_{k,l}$ is the \textit{coherence} of the $(k,l)^{\text{th}}$ sublock of $U$, defined as
\begin{equation*}
\mu_{k,l} = \max_{\substack{N_{k-1} < i \leq N_k \\ M_{l-1} < j \leq M_l}} | u_{ij} |^2.
\end{equation*}
The main point is that one can use this guarantee, along with some understanding of the local coherences $\mu_{k,l}$ to design a sampling scheme $\Omega$ that exploits the local sparsity in levels structure.  An important instance of this setup is the case of Fourier sampling with wavelets\cite{LiAdcockRIP}, in which case $\Omega$ corresponds to the frequencies sampled.  Binary sampling with the Walsh--Hadamard transform has also been considered\cite{AAHWalshWavelet}.  In both cases, designing the sampling scheme in this way to exploit the underlying structure can lead to significant benefits.\cite{AHPRBreaking,AsymptoticCS}

\section{IHT and CoSaMP in Levels}

We now introduce structured recovery algorithms for the sparsity in levels model, based on IHT and CoSaMP respectively.

\subsection{Definitions}

Fix sparsity levels $\bm{M} = (M_1,\ldots,M_r)$.  Note that any vector $x \in \mathbb{C}^N$ can be written uniquely as $x = \sum^{r}_{k=1} x_k$, where $x_k \in \mathbb{C}^N$ with $\mathrm{supp}(x_k) \subseteq \{M_{k-1}+1,\ldots,M_k \}$.  Now let $\bm{s} = (s_1,\ldots,s_r)$ be local sparsities.  For $x \in \mathbb{C}^{N}$, we write  $L_{\bm{s},\bm{M}}(x)$ for the set
\begin{equation*}
    L_{\bm{s},\bm{M}}(x) = \bigcup_{k=1}^{r} L_{s_{k}}(x_k).
\end{equation*}
In other words, this is the index set consisting, in each level $\{ M_{i-1}+1, \ldots , M_{i} \}$, of the largest absolute $s_{i}$ entries of $x$ in that level.
With this in hand, we define the \textit{hard thresholding in levels} operator $H_{\bm{s},\bm{M}} : \mathbb{C}^N \rightarrow \mathbb{C}^N$ by
\begin{equation*}
H_{\bm{s},\bm{M}}(x) = (H_{\bm{s},\bm{M}}(x)_{i})_{i=1}^{N},\qquad     H_{\bm{s},\bm{M}}(x)_{i} = \begin{cases} x_{i} &i \in L_{\bm{s},\bm{M}}(x) \\
    0 &\text{otherwise}
        \end{cases},\qquad x = (x_i)^{N}_{i=1} \in \mathbb{C}^{N}.  
\end{equation*}
That is, $H_{\bm{s},\bm{M}}(x)$ is the vector consisting of the largest $(\bm{s},\bm{M})$ entries of $x$ with all other entries set to zero.

The levels versions of the classical IHT and CoSaMP algorithms now follow simply by replacing the thresholding steps by the above levels versions.  Specifically, \textit{IHT in Levels (IHTL)} is defined by

\begin{tcolorbox}
Function $\mathrm{IHTL}(A,y,\bm{s},\bm{M})$ \\
\noindent \textbf{Inputs:} $A \in \mathbb{C}^{m\times N}$, $y \in \mathbb{C}^m$, local sparsities $\bm{s}$, sparsity levels $\bm{M}$
\\
\noindent \textbf{Initialization:} $x^{(0)} \in \mathbb{C}^N$ (e.g.\ $x^{(0)} = 0$) \\
\noindent \textbf{Iterate:} Until some stopping criterion is met at $n = \overline{n}$, set \\
\begin{equation*}
    x^{(n+1)}= H_{\bm{s},\bm{M}}(x^{(n)} + A^{*}(y - Ax^{(n)}))
\end{equation*}

\noindent \textbf{Output:} $\hat{x} = x^{(\overline{n})}$

\end{tcolorbox}

and \textit{CoSaMP in Levels (CoSaMPL)} is defined by

\begin{tcolorbox}

Function $\mathrm{CoSaMPL}(A,y,\bm{s},\bm{M})$ \\
\noindent \textbf{Inputs:} $A \in \mathbb{C}^{m\times N}$, $y \in \mathbb{C}^m$, local sparsities $\bm{s}$, sparsity levels $\bm{M}$
\\
\noindent \textbf{Initialization:} $x^{(0)} \in \mathbb{C}^N$ (e.g.\ $x^{(0)} = 0$) \\
\noindent \textbf{Iterate:} Until some stopping criterion is met at $n = \overline{n}$, set 
\begin{align*}
    U^{(n+1)} &= \text{supp}(x^{(n+1)}) \cup L_{2\bm{s},\bm{M}}(A^{*}(y-Ax^{(n)})) \\
    u^{(n+1)} & \in \argmin_{z \in \mathbb{C}^N} \{ \Vert y -Az \Vert_{2} \: : \: \text{supp}(z) \subset U^{(n+1)} \} \\
    x^{(n+1)} &= H_{\bm{s},\bm{M}}(u^{(n+1)})
\end{align*}

\noindent \textbf{Output:} $\hat{x} = x^{(\overline{n})}$

\end{tcolorbox}

\subsection{Experiments}
\label{sec:Experiments}

We now present a series of numerical experiments.  Our aim is to demonstrate the benefits that the structure-promoting IHTL and CoSaMPL algorithms bring for sparse in levels vectors over the standard IHT and CoSaMP algorithms.  To do this, we consider phase transition plots.

For each fixed total sparsity $s$ and number of measurements $m$ we generate an $(\bm{s},\bm{M})$-sparse in levels vector $x$ of length $N = 128$ with random support and unit normal random entries.  Note that, as we shall see below, the local sparsities $\bm{s}$ are related in some way to the total sparsity $s$.  We then compute its reconstruction $\hat{x}$ using either IHT, IHTL, CoSaMP or CoSaMPL and calculate the relative error $\Vert x - \hat{x} \Vert_{\ell^2} / \Vert x \Vert_{\ell^2}$.  This is repeated for 50 trials, and the empirical success probability calculated.  A recovery is successful if $\Vert x - \hat{x} \Vert_{\ell^2} / \Vert x \Vert_{\ell^2} < 10^{-2}$.  The measurement matrix $A$ is a Gaussian random matrix (independent, normally distributed entries with mean zero and variance $1/\sqrt{m}$).
Each algorithm is halted when either the relative difference between $x^{(n+1)}$ and $x^{(n)}$ is less than a tolerance $10^{-4}$ or if $n$ exceeds 1000 iterations. Moreover, we choose the initialization $x^{(0)} = 0$.



For IHT and IHTL, in order to obtain better performance we apply a rescaling by the factor $\sqrt{m/N}$ and compute $\hat{x} = \mathrm{IHT}(\sqrt{m/N} A , \sqrt{m/N} y , s)$. Note that rescaling $A$ corresponds to changing the step size of the Landweber iteration before thresholding. We observe the convergence of IHT is guaranteed by the sufficient condition $\|A\|_2 < 1$.\cite{BlumensathDavies2008}. Using random matrix theory,\cite{vershynin2018high} it is possible to see that in the cases considered here $\|A\|_2 \lesssim \sqrt{N/m}$ with high probability, which leads to choosing the scaling factor $\sqrt{m/N}$. Other than this, our results consider vanilla versions of all algorithms: our goal is to examine the benefits of sparsity in levels over classical sparsity, rather than the intrinsic performance of the decoders themselves.  Notice, as a general rule, that CoSaMP outperforms IHT. 

\begin{figure} [ht]
\begin{center}
\begin{tabular}{ccc} 
\includegraphics[height=\figthree]{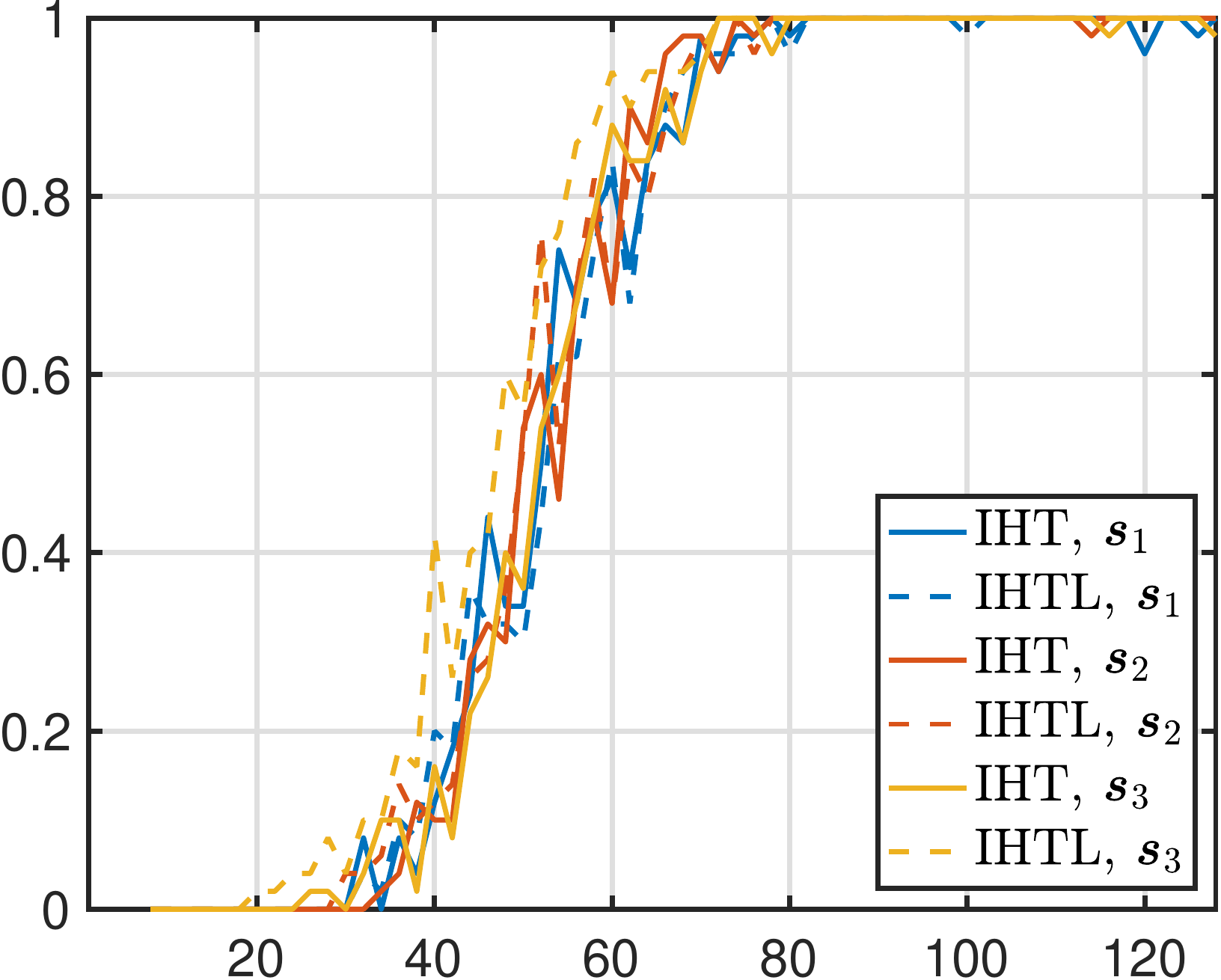} &
\includegraphics[height=\figthree]{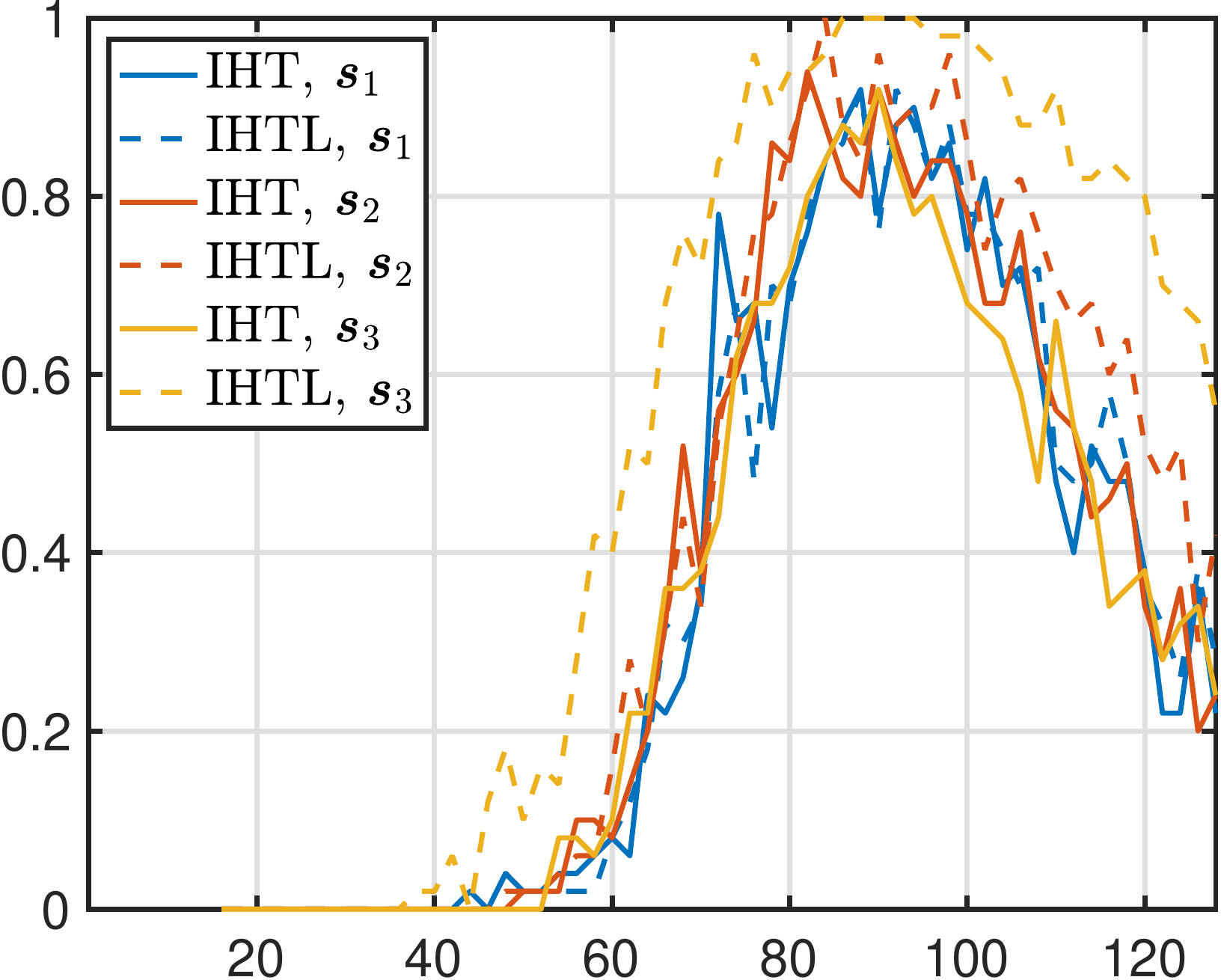} &
\includegraphics[height=\figthree]{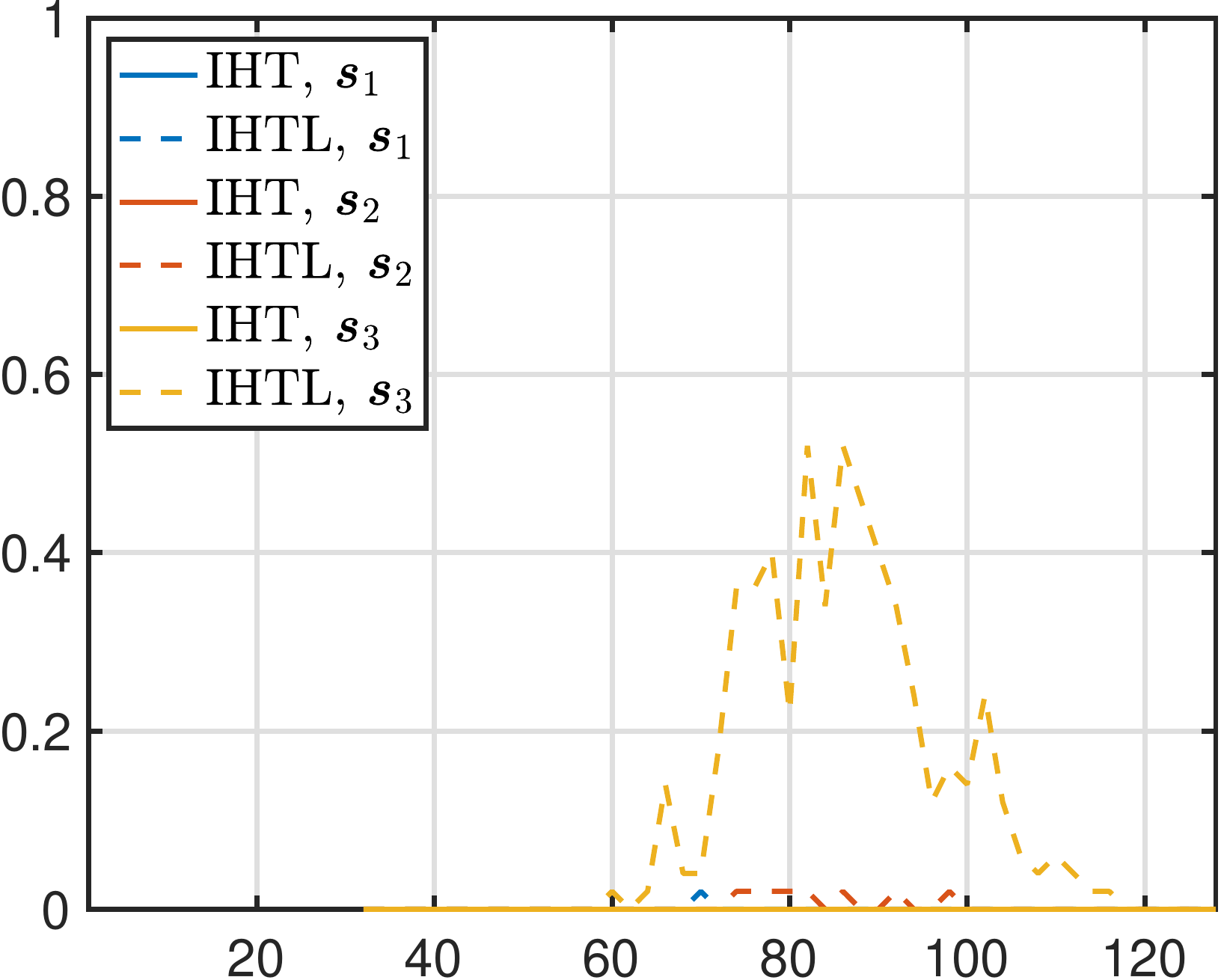} 
\\
\includegraphics[height=\figthree]{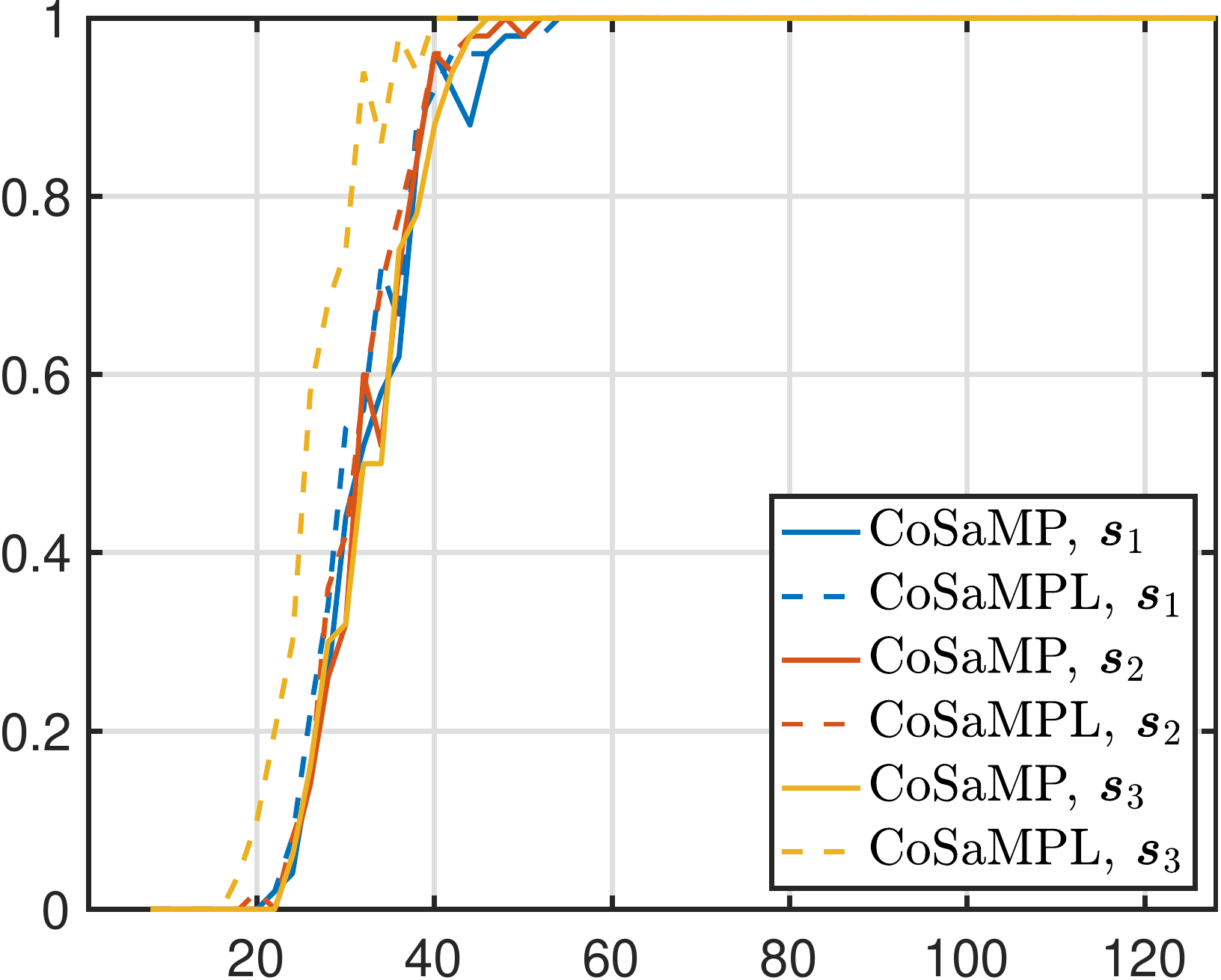} &
\includegraphics[height=\figthree]{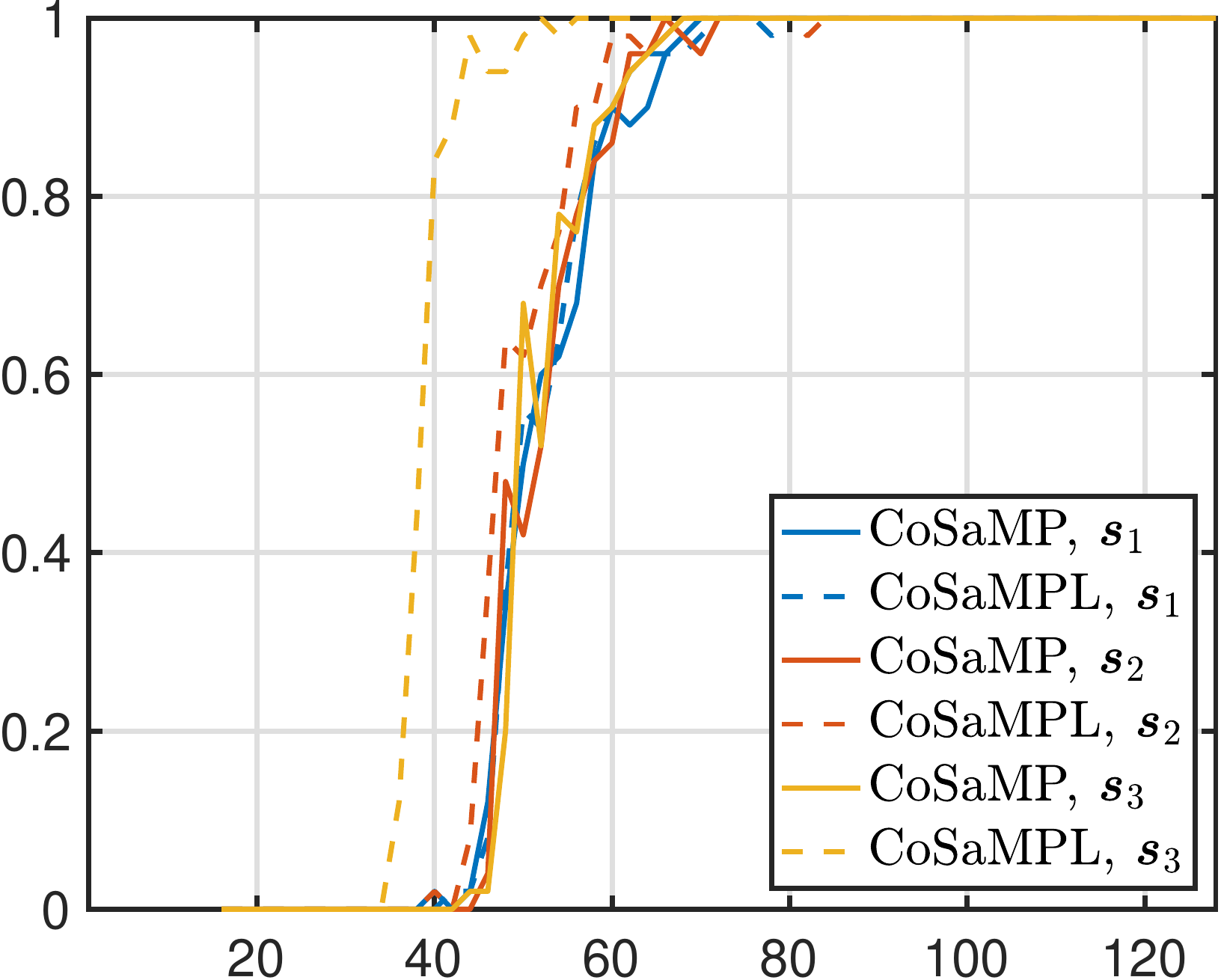} &
\includegraphics[height=\figthree]{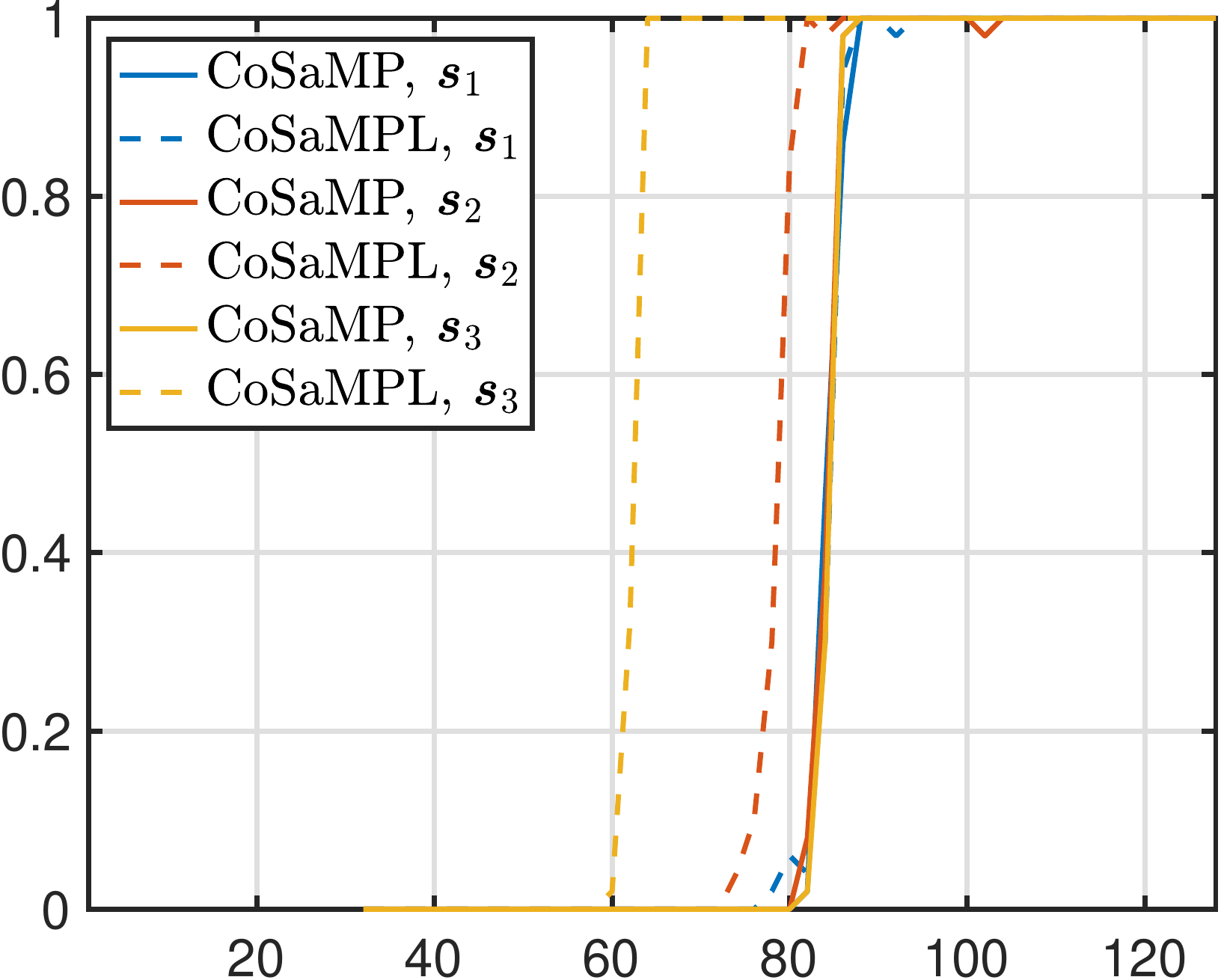} 
\\
$s = 8$ & $s = 16$ & $s = 32$
\end{tabular}
\end{center}
\caption[example] 
{ \label{fig:2levelunif} 
Horizontal phase transition line showing success probability versus $m$ for various fixed total sparsities $s$.  Two level sparsity with $\bm{M} = (N/2,N)$.  The local sparsities are $\bm{s}_1 = (s/2,s/2)$, $\bm{s}_2 = (3s/4,s/4)$ and $\bm{s}_3 = (s,0)$.  The top row considers IHT and IHTL.  The bottom row considers CoSaMP and CoSaMPL.
}
\end{figure}

Our first experiment, shown in Figure \ref{fig:2levelunif}, considers the two-level case.  The levels are chosen of equal size $N/2$, where $N = 128$, and we use various different local sparsities. Namely, we consider $\bm{s}_1 = (s/2, s/2)$, $\bm{s}_2 = (3s/4, s/4)$, $\bm{s}_3 = (s,0)$ and $s =8, 16, 32$.  As expected, when the local sparsities are $\bm{s}_1 =  (s/2,s/2)$ there is no benefit to either IHTL or CoSaMPL over IHT or CoSaMP.  However, as the local sparsities become more unbalanced one starts to see benefits.  In the extreme case $\bm{s}_3 = (s,0)$, CoSaMPL with $s = 32$ achieves successful recovery with probability one from roughly 65 measurements, while CoSaMP requires roughly 90 measurements.

\begin{figure} [ht]
\begin{center}
\begin{tabular}{ccc} 
\includegraphics[height=\figthree]{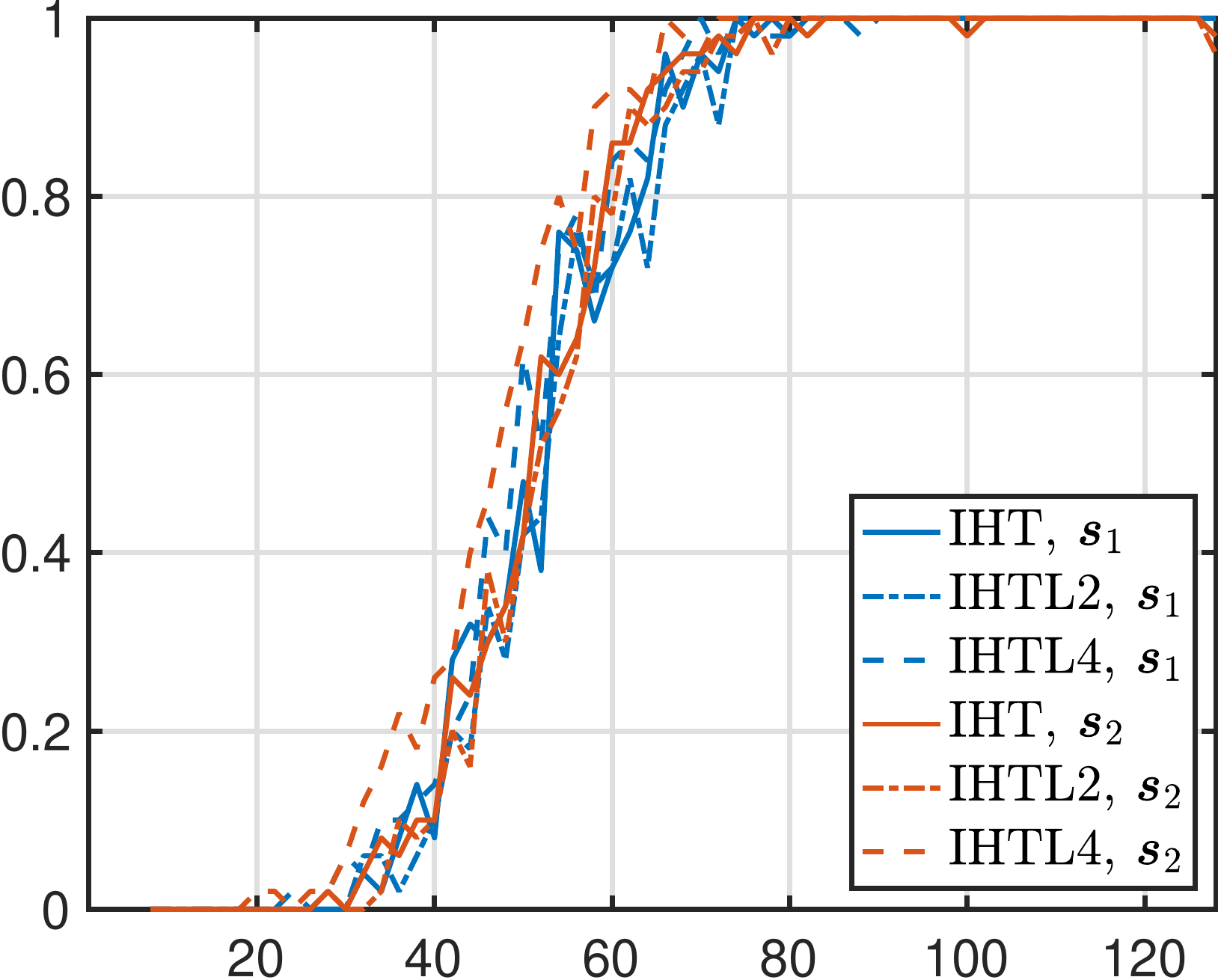} &
\includegraphics[height=\figthree]{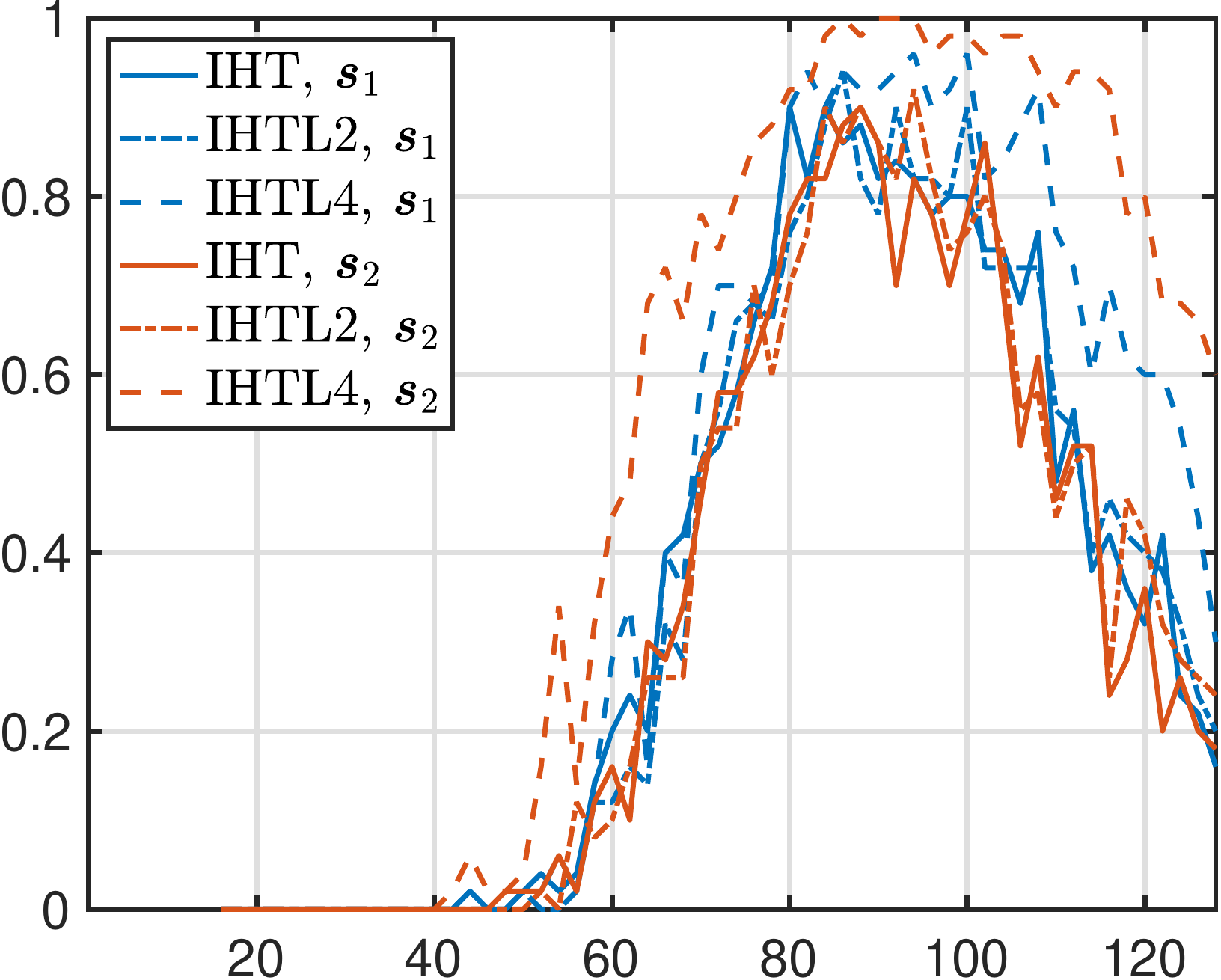} &
\includegraphics[height=\figthree]{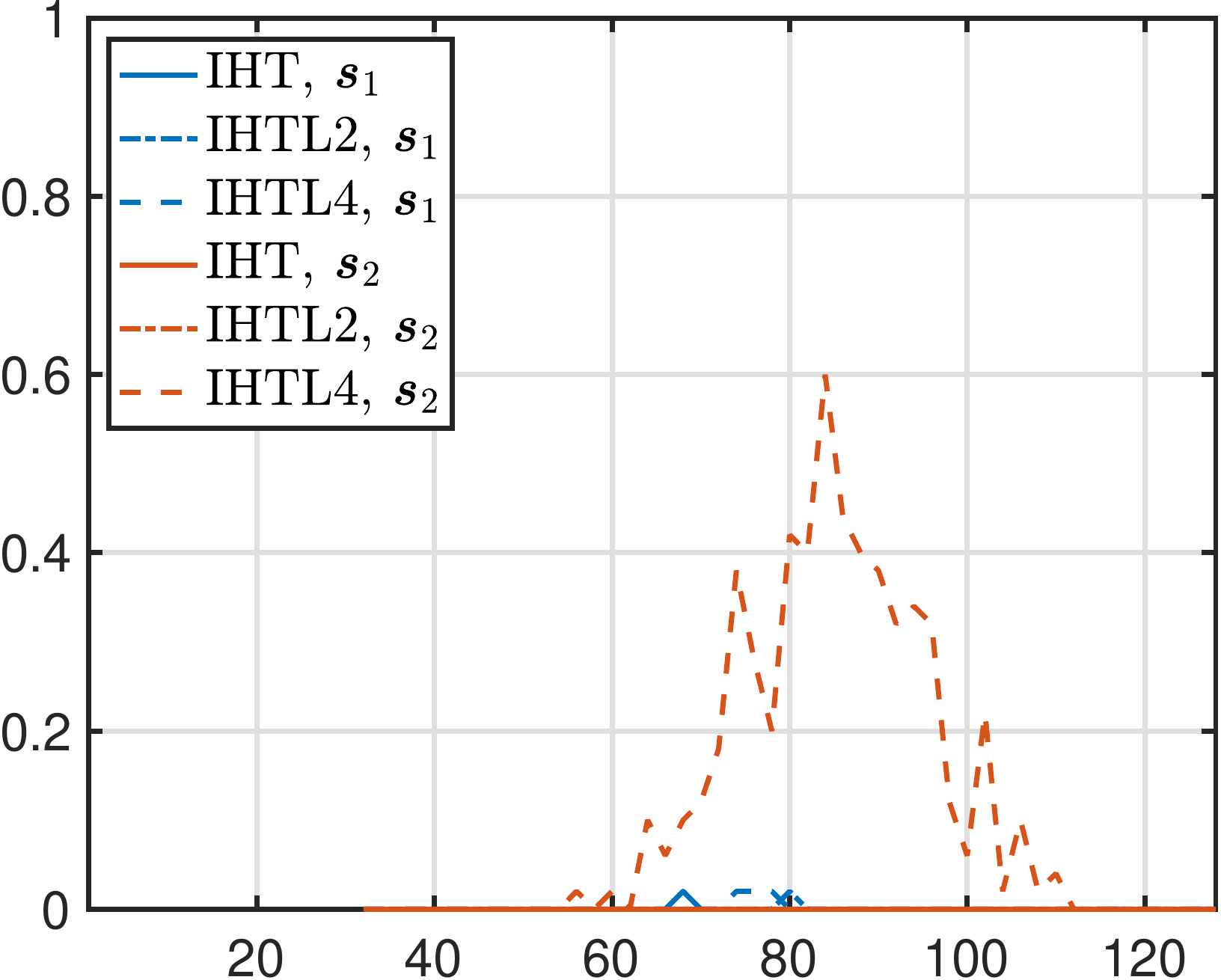} 
\\
\includegraphics[height=\figthree]{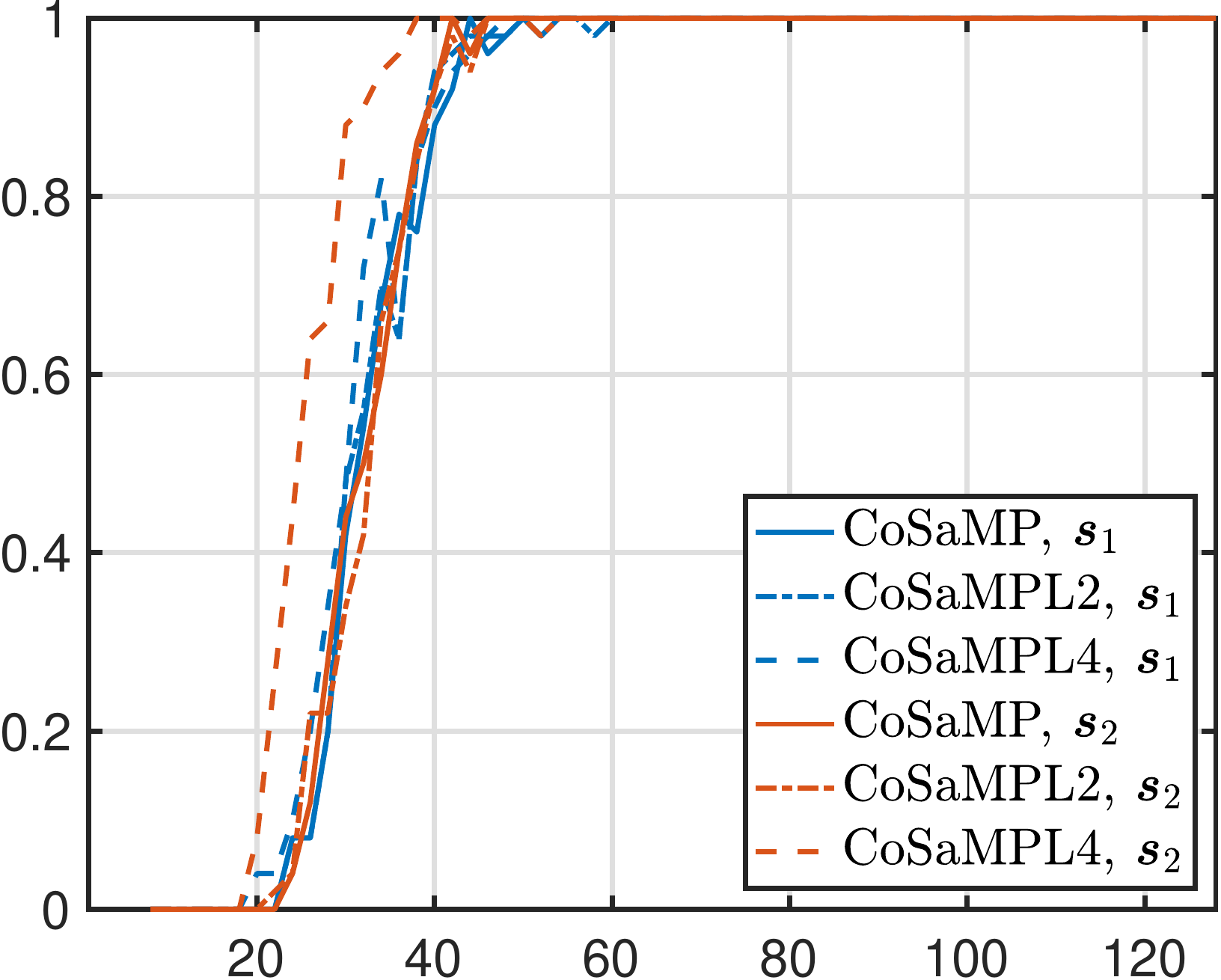} &
\includegraphics[height=\figthree]{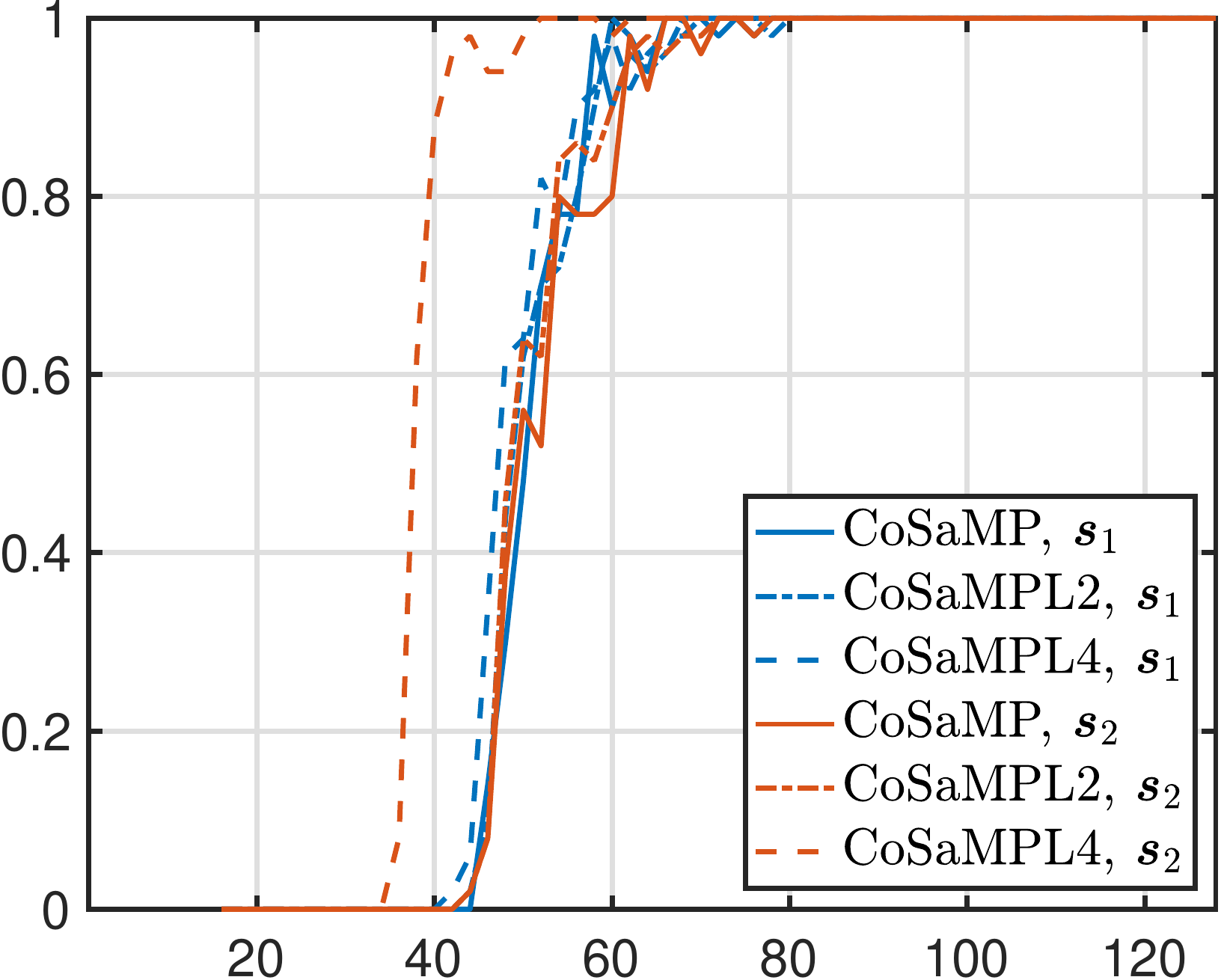} &
\includegraphics[height=\figthree]{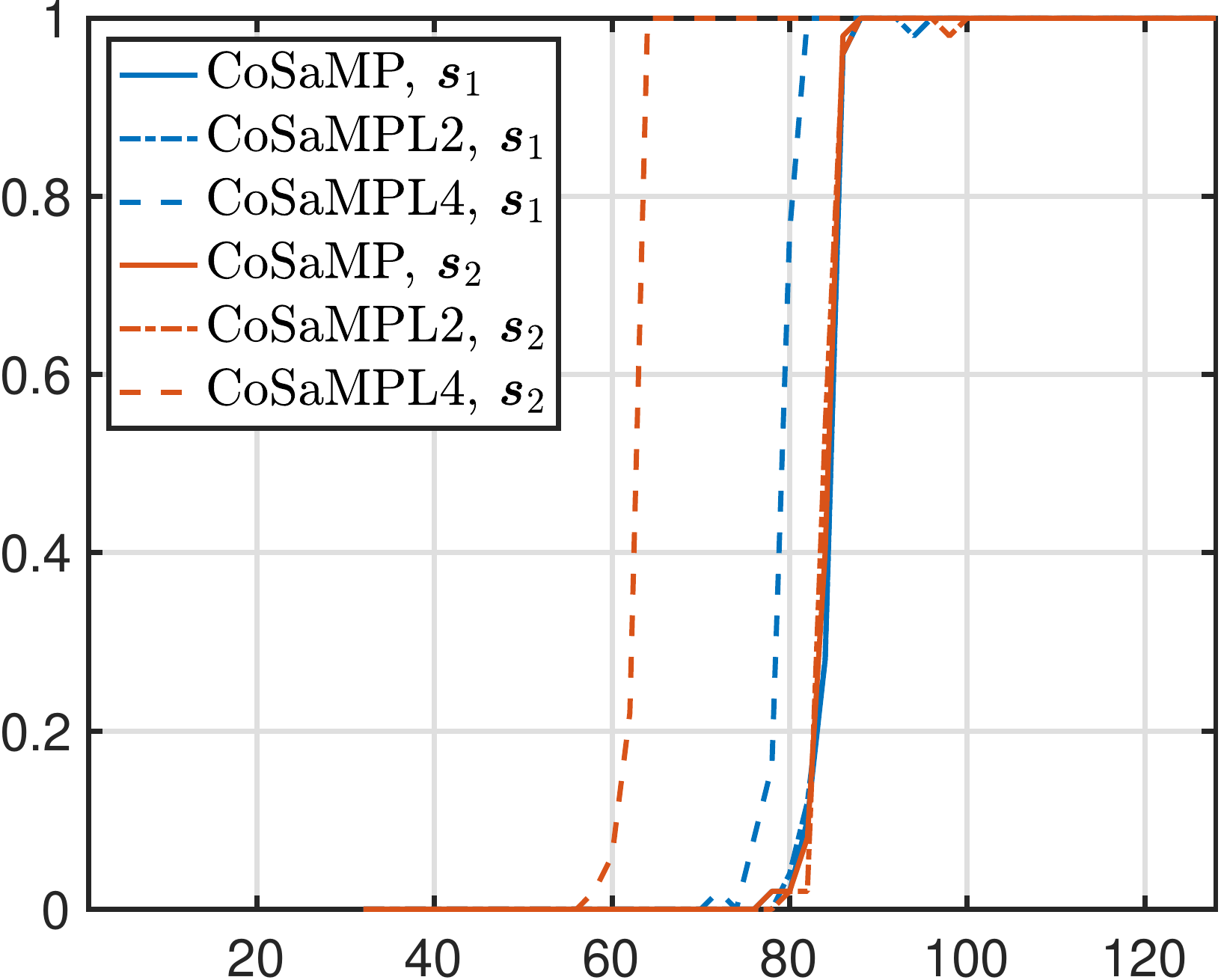} 
\\
$s = 8$ & $s = 16$ & $s = 32$
\end{tabular}
\end{center}
\caption[example] 
{ \label{fig:4levelunif} 
Horizontal phase transition line showing success probability versus $m$ for various fixed total sparsities $s$.  Four level sparsity with $\bm{M} = (N/4,N/2,3N/4,N)$.  The local sparsities are $\bm{s}_1 = (3 s/8 , s/8 , 3 s /8 , s/8)$ and $\bm{s}_2 = (s/2 , 0 , s/2 , 0)$.  
The top row considers IHT and IHTL.  The bottom row considers CoSaMP and CoSaMPL.  In the levels case we consider two-level algorithms (IHTL2 and CoSaMPL2) based on $\bm{M} = (N/2,N)$ and $\bm{s} = (s/2,s/2)$ and four-level algorithms (IHTL4 and CoSaMPL4) based on $\bm{M} = (N/4,N/2,3N/4,N)$ and $\bm{s} = \bm{s}_1$ or $\bm{s} = \bm{s}_2$.
}
\end{figure}

\begin{figure} [ht]
\begin{center}
\begin{tabular}{cc} 
\includegraphics[height=\figtwo]{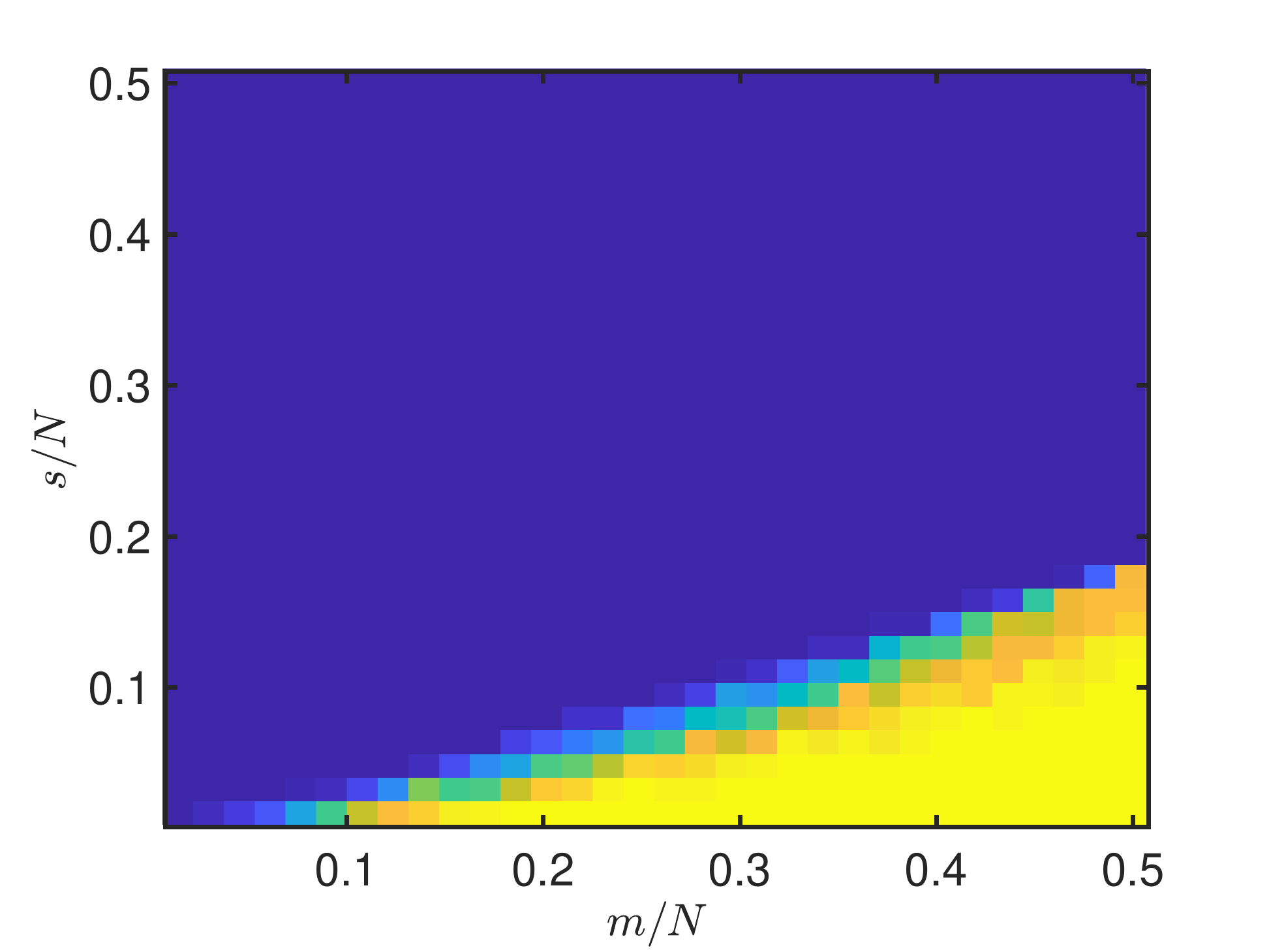}
&
\includegraphics[height=\figtwo]{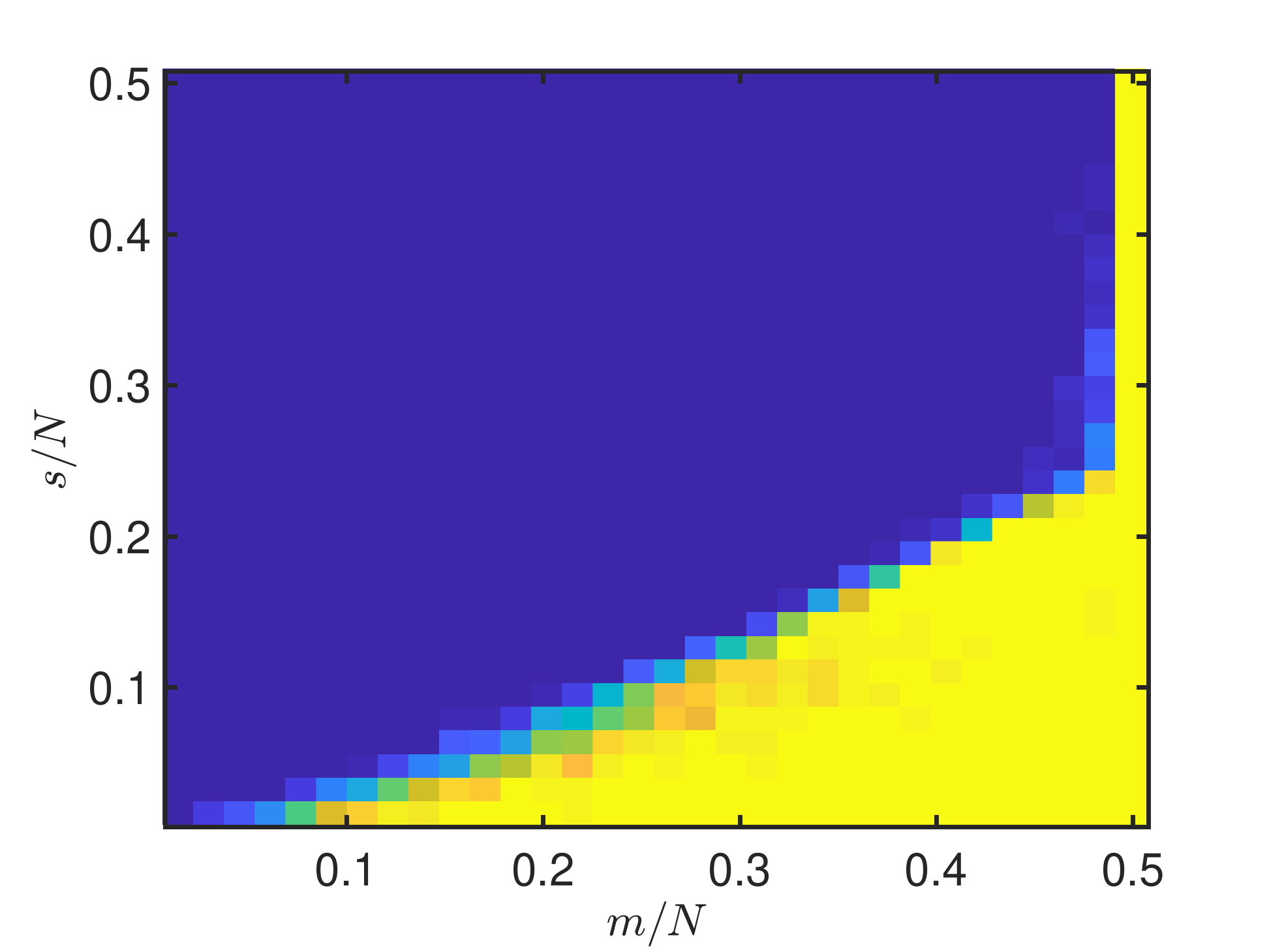} 
\\
CoSaMP & CoSaMPL
\end{tabular}
\end{center}
\caption[example] 
{ \label{fig:4leveluniffullPT} 
Full phase transition for CoSaMP and CoSaMPL.  Four level sparsity with $\bm{M} = (N/4,N/2,3N/4,N)$ and $\bm{s} = (s/2,0,s/2,0)$.  We restrict $s$ to $s \leq N/2$ since when $s > N/2$ the local sparsity in the first and third levels exceed the size of the levels. Yellow and blue pixels correspond to success probabilities 1 and 0 respectively.

}
\end{figure}

In Figure \ref{fig:4levelunif} we consider four levels, again equally-sized, with $\bm{M} = (N/4,N/2,3N/4,N)$.  We compare the IHT and CoSaMP algorithms with the IHTL and CoSaMPL algorithms.  The local sparsities $\bm{s}$ take the form $\bm{s} = (a s , b s , as , b s)$, where $a + b = 1/2$, for different values of $a$ and $b$.  In this experiment we use two versions of IHTL and CoSaMPL, based on two levels or four levels.  In the two levels algorithms we use the values $(s/2,s/2)$ and $(N/2,N)$ for the decoders.  This is because a vector that is $(\bm{s},\bm{M})$-sparse with the local sparsities $\bm{s} = (a s , b s , as , b s)$ is also $((s/2,s/2),(N/2,N))$-sparse.  As one would expect, the 2-level algorithms give no benefit over the original (1-level) algorithms.  Yet, as in the previous experiment, we see a significant benefit from the 4-level algorithms.  This figure considers several fixed values of the total sparsity $s$.  In Figure \ref{fig:4leveluniffullPT} we give the full phase transitions for CoSaMP and CoSaMPL.  We notice the significantly improved transition curve.  Note that CoSaMPL achieves probability one recovery when $m/N = 1/2$ for any $s$.  The reason for this is that in this case the levels either have no nonzero entries, or all their entries are nonzero.  CoSaMPL exploits this structure, however CoSaMP cannot.

\begin{figure} [ht]
\begin{center}
\begin{tabular}{ccc} 
\includegraphics[height=\figthreealt]{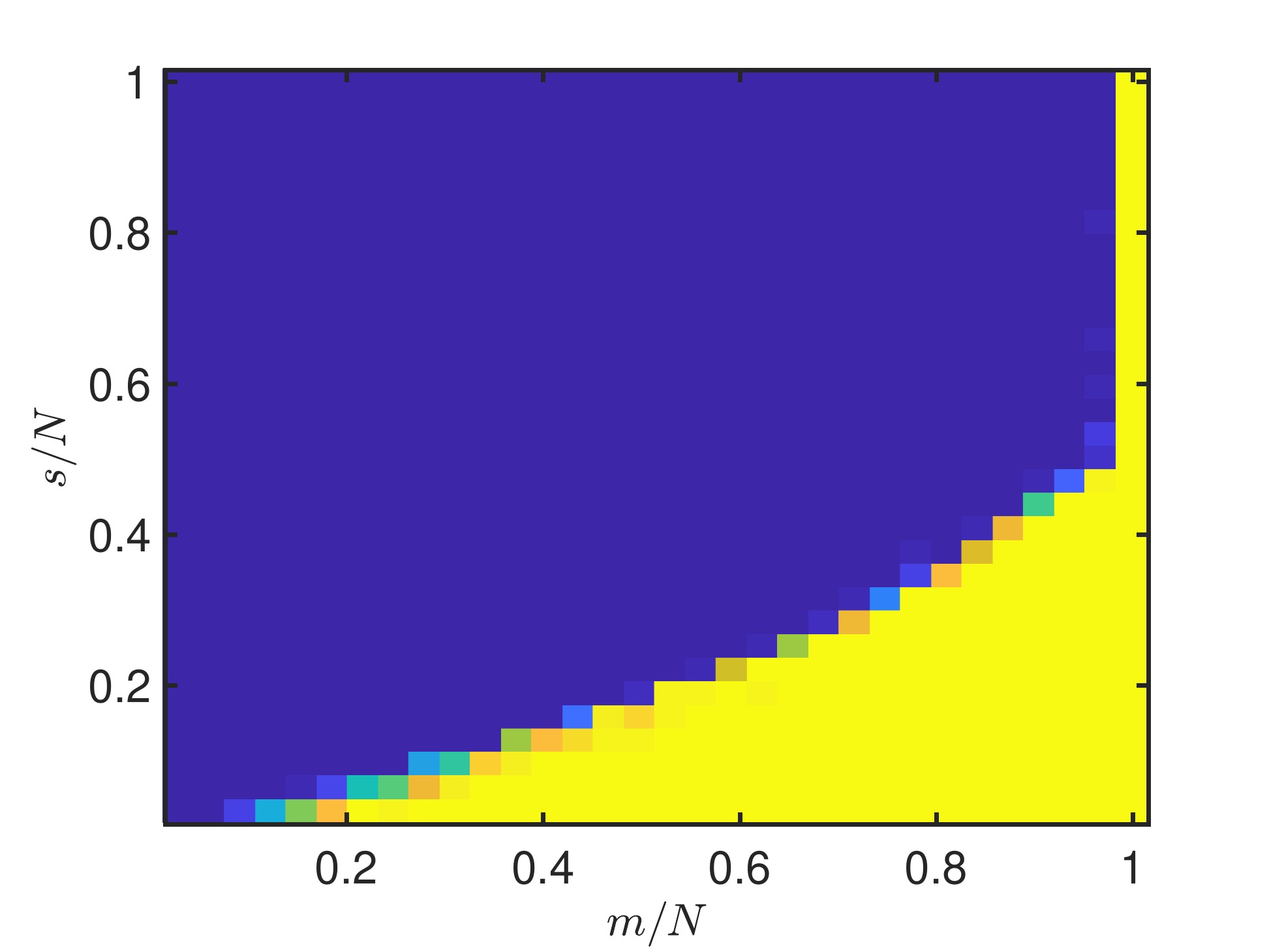} &
\includegraphics[height=\figthreealt]{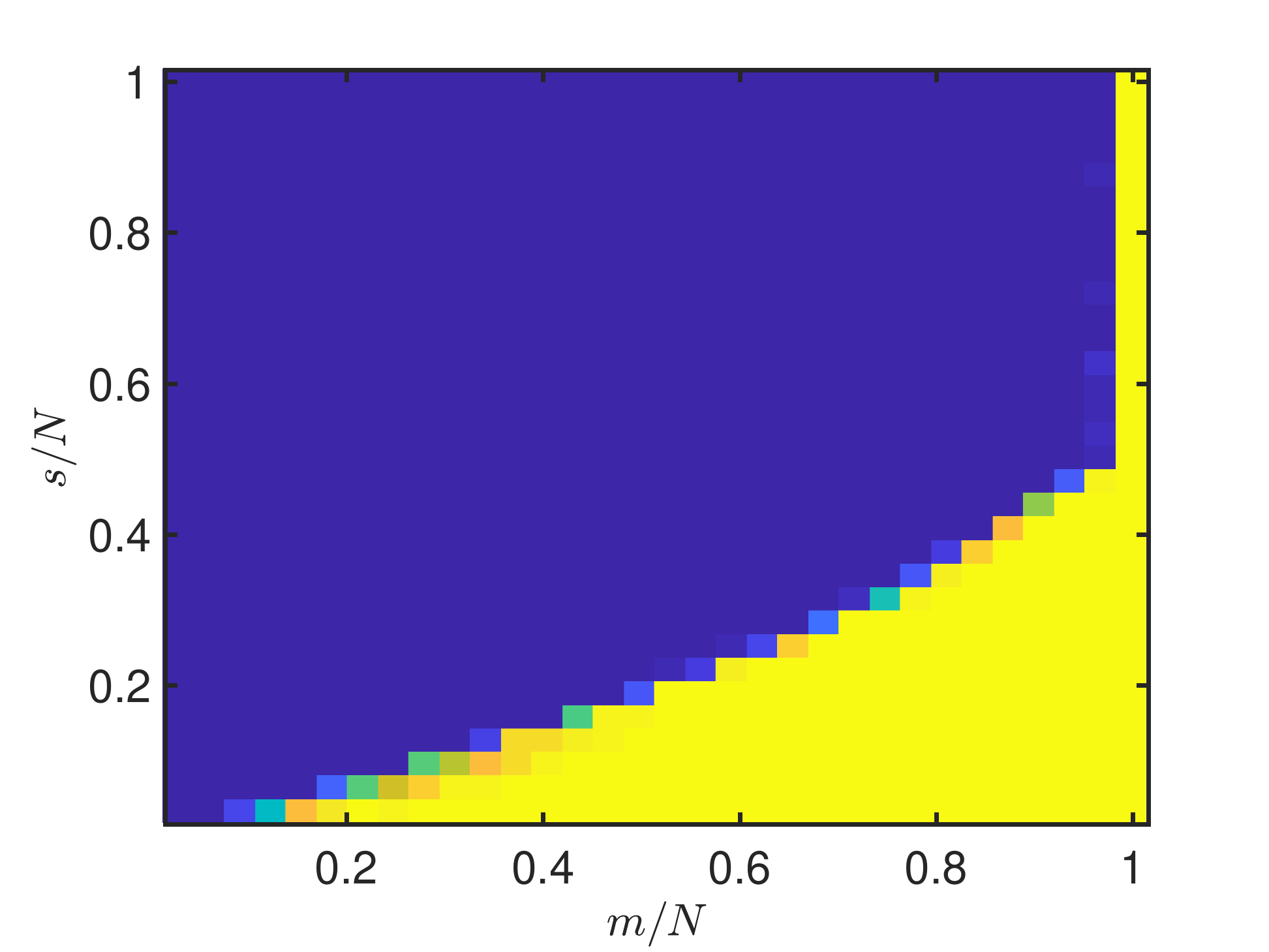} &
\includegraphics[height=\figthreealt]{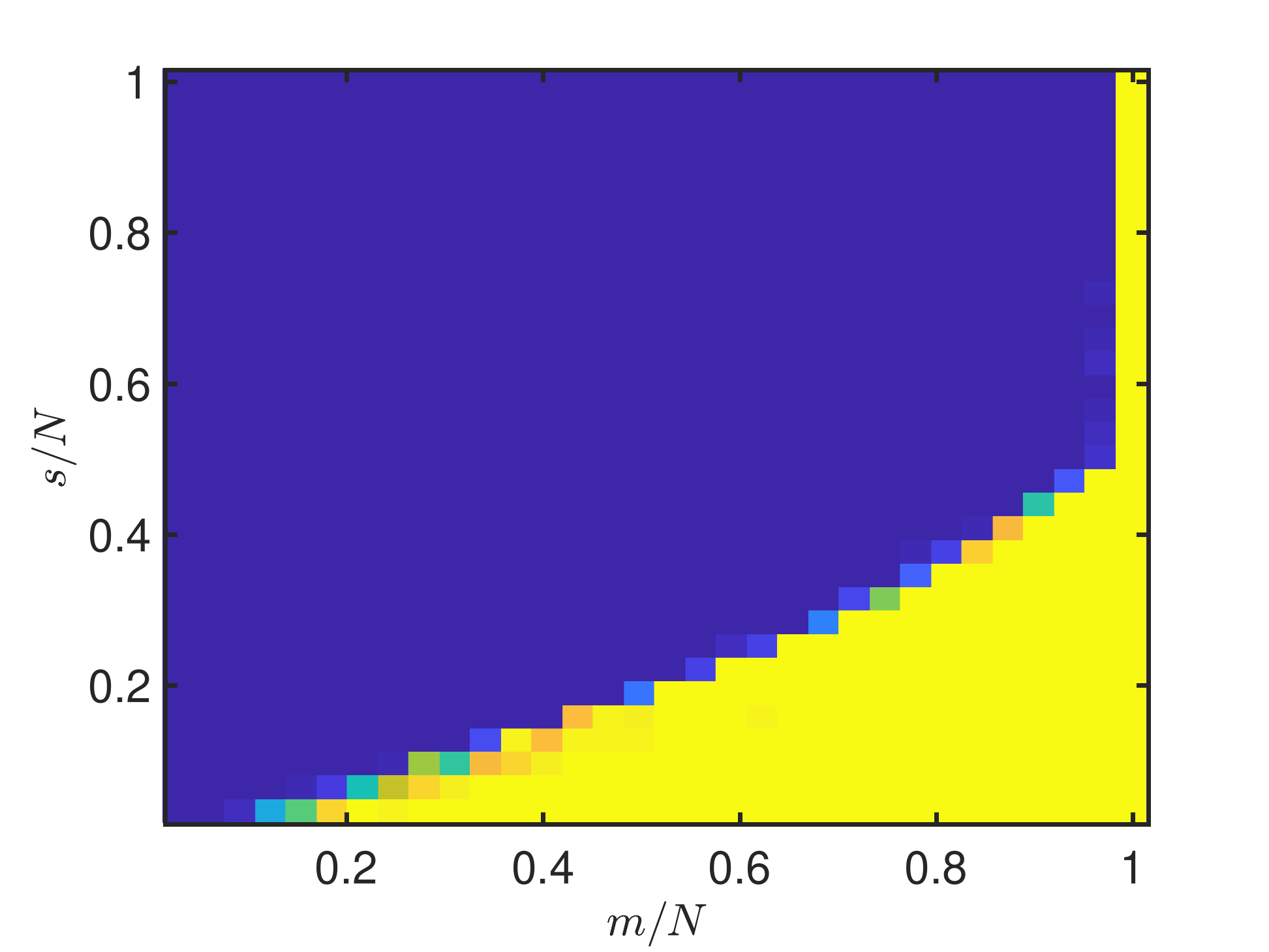} 
\\
\includegraphics[height=\figthreealt]{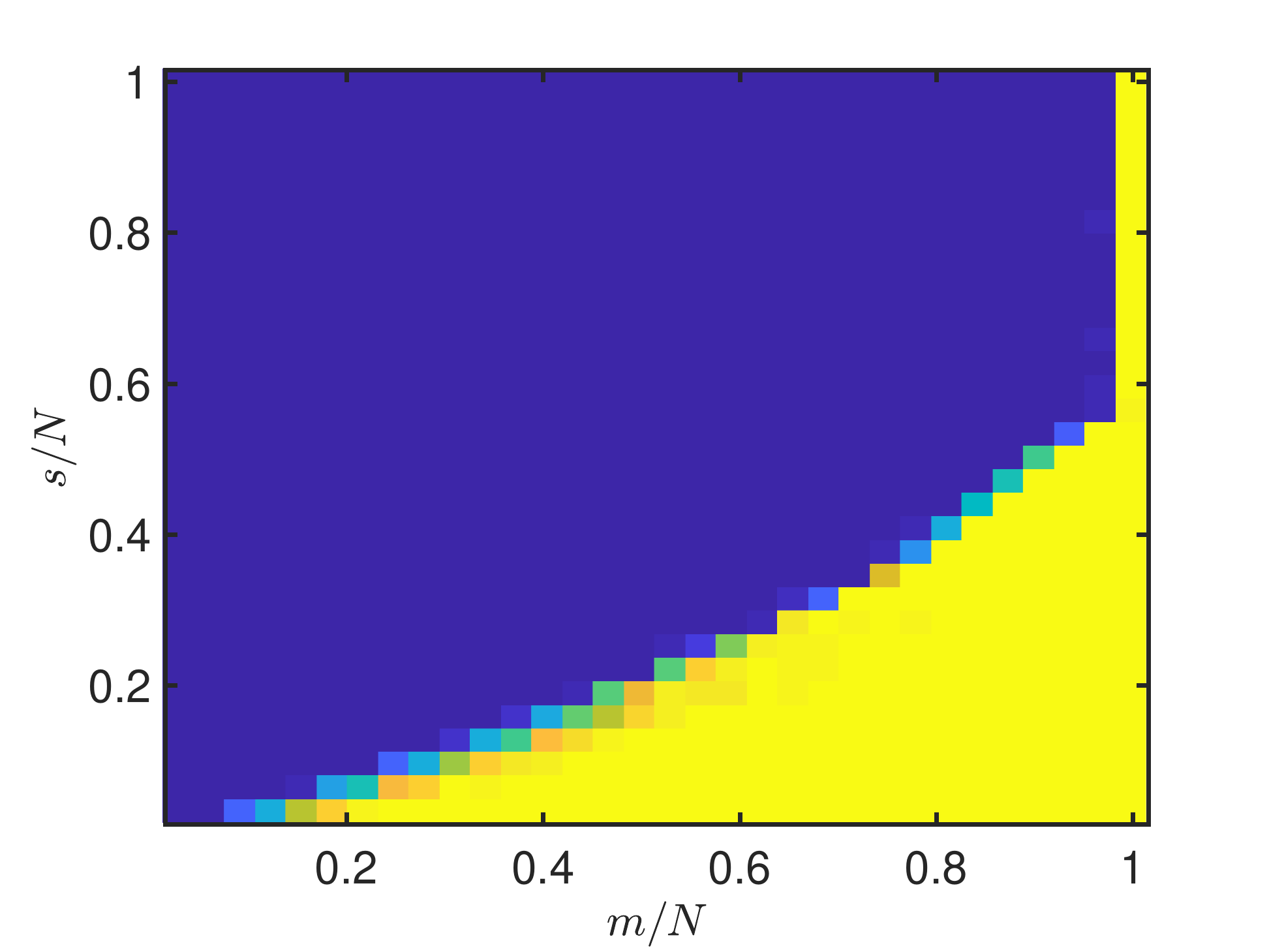}  &
\includegraphics[height=\figthreealt]{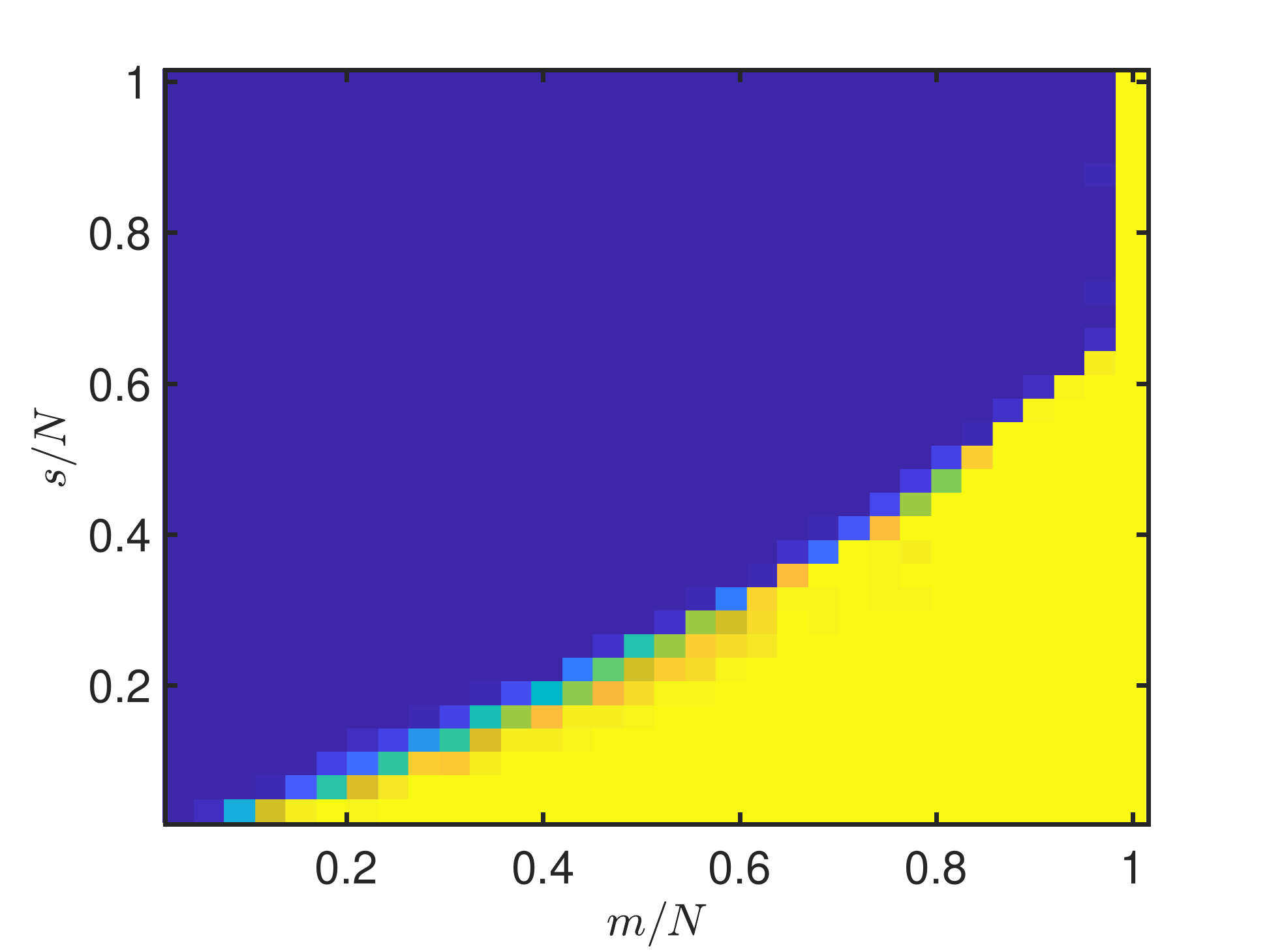} &
\includegraphics[height=\figthreealt]{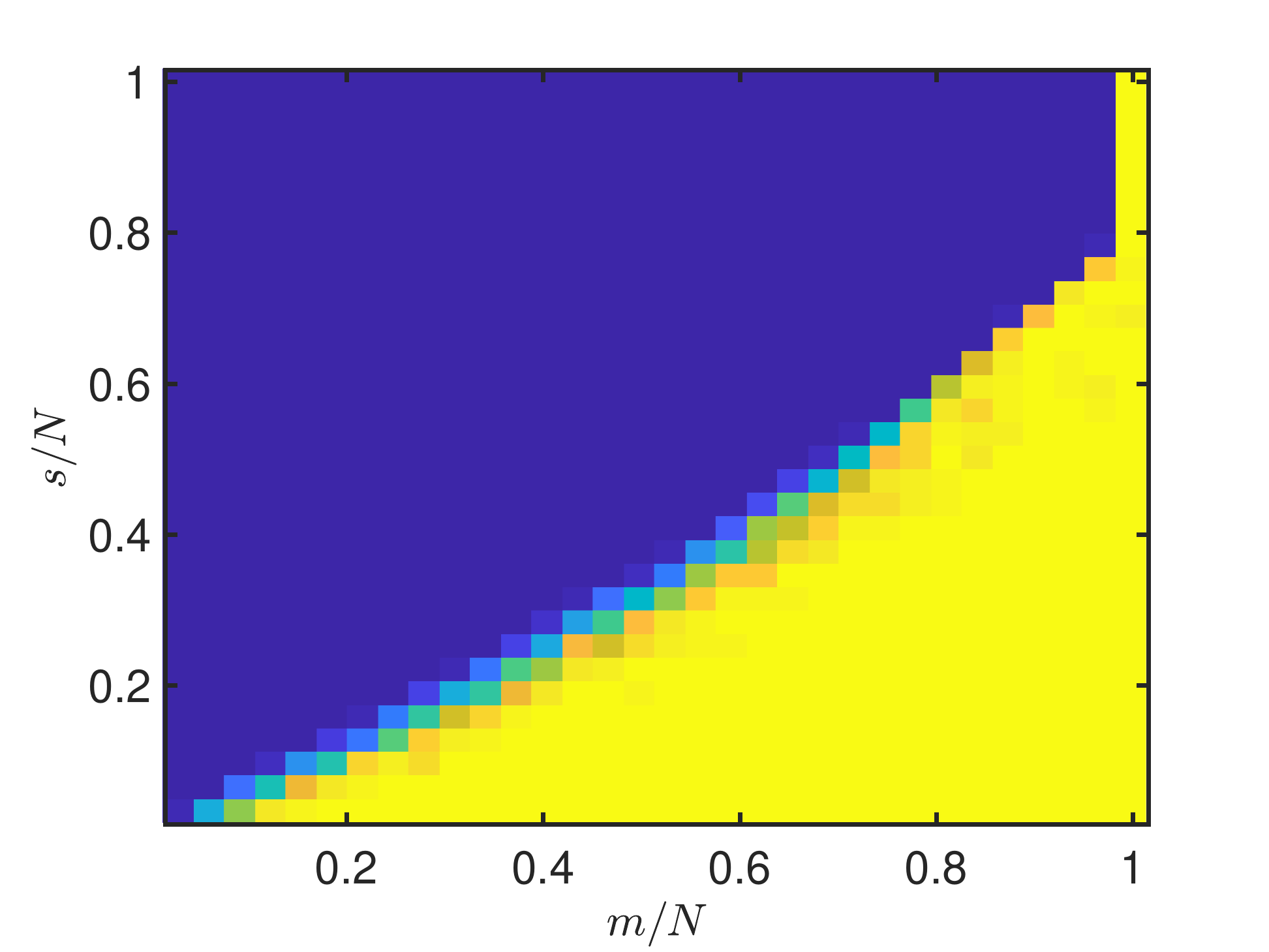} 
\\
$a = 1/4$ & $a = 1/2$ & $a = 3/4$
\end{tabular}
\end{center}
\caption[example] 
{ \label{fig:2levelsaturate} 
Phase transitions for CoSaMP (top row) and CoSaMPL (bottom row).  Two-level sparsity with levels $\bm{M} = (a s , N)$ and local sparsities $\bm{s} = (as,(1-a) s)$ for different values of $0 < a < 1$.
}
\end{figure}

Next, in Figure \ref{fig:2levelsaturate} we consider a rather different setup.  Here, given $s$ we consider two-level sparsity with levels taking the form $\bm{M} = (M_1,M_2) = (a s , N)$, for some $0 < a < 1$.  In other words, the $s$-sparse vectors that are generated are nonzero in their first $a s$ entries, with the remaining $(1-a) s$ entries being arbitrarily located among the indices $\{ a s + 1,\ldots,N\}$.  For succinctness we consider only CoSaMP in this experiment.

The purpose of this experiment is to model a typical scenario in compressed sensing with wavelet sparsifying transforms, where the first $a s$ wavelet coefficients are `saturated', i.e.\ all nonzero.  We discuss this further in the next section.  For now, however, we simply notice the benefits of CoSaMPL over standard CoSaMP.  For example, if $s = 32$ and $3/4$ of the coefficients are saturated, CoSaMP requires 33\% more measurements to achieve successful recovery.


\subsection{Function approximation via compressed sensing}


Finally, we test the proposed IHTL and CoSaMPL algorithms in the context of function approximation via compressed sensing\cite{BASBMKRCSwavelet}. We aim to approximate a function $f : [0,1] \to \mathbb{R}$. In this context, it is convenient to adopt the terminology of \emph{decoders} and \emph{encoders}.  An encoder is a linear mapping $\mathcal{E}_m : L^2([0,1]) \to \mathbb{C}^m$ corresponding to the measurement phase. A decoder is a mapping $\mathcal{D}_m : \mathbb{C}^m \to L^2([0,1])$ and corresponds to the recovery phase. We focus on the class of piecewise $\alpha$-H\"older functions, defined as the set of functions with a finite number of discontinuities and $\alpha$-H\"older continuous over the intervals of smoothness. 

We compute an approximation $\tilde{f}_m = \mathcal{D}_m(\mathcal{E}_m(f))$ of $f$. Our goal is to find encoder-decoder pairs such that the approximation error $\|f - \tilde{f}_m\|_{L^2}$ decays at a rate as close as possible to the theoretically optimal $O(m^{-\alpha})$.\cite{mallat09wavelet, devore1993wavelet, BASBMKRCSwavelet} Multilevel Fourier sampling sampling strategies have  been recently showed to achieve a near-optimal approximation rate $O(\log^\gamma(m) / m^{\alpha})$ with $\gamma = 13/4 +  \delta$ for any $0 < \delta < 1$, when combined with Daubechies' wavelet approximation and with a decoder based on $\ell^1$ minimization\cite{BASBMKRCSwavelet} (more precisely, the so-called weighted square-root LASSO decoder\cite{ABBCorrecting}). This near-optimal result heavily relies on the sparsity in levels structure. In fact, the specific pattern of sparsities in levels of wavelet coefficients that leads to the optimal approximation error rate is captured by devising an \emph{ad hoc} multilevel sampling strategy which saturates the lower frequencies bands and increasingly subsamples the higher ones.

The aim of the numerical experiments performed in this section is to investigate the different role played by structure in the encoder and in the decoder. Indeed, the existence of optimal and near-optimal encoder-decoder pairs has only been proved where the structure in levels is exploited in the encoder but not in the decoder. For this reason, we consider a structure-agnostic and a structure-promoting encoder based on random Gaussian sampling and on multilevel Fourier sampling, respectively. Let $\{\phi_n\}_{n \in \mathbb{N}}$ be the Haar wavelet basis and fix a truncation level $N$ and assume, for the sake of simplicity, $N$ and $m$ to be powers of 2 and $m \geq 2$. The two encoders are defined as follows:
\begin{description}
\item [Gaussian encoder (structure agnostic)] The action of $\mathcal{E}_m$ ``shuffles'' the wavelet coefficients of $f$ via random Gaussian sampling. Namely, $\mathcal{E}_m (f) = A (\langle f, \phi_n\rangle)_{n = 1}^N$, where $A$ is an $m \times N$ matrix whose entries are i.i.d.\ centered random Gaussian variables with mean $0$ and variance $1/\sqrt{m}$. 

\item [Fourier encoder (structure promoting)] It has the form $\mathcal{E}_m(f) = (\hat{f}(2\pi k_i ))_{i = 1}^m$, where $\hat{f}(\omega) = \int_{-\infty}^{+\infty} f(t) e^{-i\omega t} \text{d}t$ is the Fourier transform of $f$. The frequencies indices $(k_i)_{i = 1}^m \subseteq \mathbb{Z}$ are sampled according to an $(\bm{m}, \bm{N})$-multilevel random sampling scheme (see Section~\ref{sec:RIPL}), roughly defined as follows. The first $m/2$ samples are used to saturate the lowest dyadic frequency bands. The remaining $m/2$ samples are evenly divided among the higher dyadic frequency bands via  subsampling.\footnote{More precisely, the set $\mathbb{Z}$ of frequency indices  is partitioned into dyadic bands 
$$
B_1 = \{0,1\}, \quad B_k = \{-2^k + 1,\ldots, -2^{k-1}\} \cup \{2^{k-1} + 1,\ldots, 2^{k}\}, \quad \text{for every } k \in \mathbb{N}.
$$
Ordering the integers in $\mathbb{Z}$ as $0,1,-1,2,-2\ldots$, and considering the first $r$ frequency bands, one obtains the vector $\bm{N} = (N_1, \ldots, N_r)$ of sampling levels defined by $N_k = 2^k$, for $k = 1,\ldots,r$. The Fourier encoder saturates the first $\tilde{r} = \log_2(m/2)$ frequency bands, i.e., $m_k = N_k - N_{k-1}$ for $k = 1, \ldots, \tilde{r}$ with $N_0 = 0$. In the higher bands $\tilde{r} < k \leq r$, the local numbers of measurements are defined as
$$
m_k = 2\left\lfloor \frac{m}{4(r-\tilde{r})}\right\rfloor, \quad k = \tilde{r} +1, \ldots, r-1,
$$ 
where, in the last frequency band, we let $m_r = m-(m_1 +\cdots + m_{r-1})$ in order to reach a total budget of exactly $m$ measurements. In order to enforce symmetry, for every $k > \tilde{r}$, we pick $m_k/2$ samples uniformly at random from the $k^{\text{th}}$ frequency semiband $B_k \cap \mathbb{N}$ and we choose frequencies in the opposite semiband in a symmetric way.\cite{BASBMKRCSwavelet}}

\end{description}

We test these two encoders when combined with five possible decoders. Namely, \emph{Basis Pursuit (BP)} (i.e., \eqref{QCBP} with $\eta = 0$), IHT, CoSaMP, IHTL, and CoSaMPL. Of course, the first three decoders are structure agnostic and the last two ones are structure promoting. Since we want all the decoders to depend on $m$ only, we fix a relation between $m$ and their input parameters ($s$ and $(\bm{s},\bm{M}))$, respectively). This leads to the definition of an auxiliary parameter $C > 0$ such that
\begin{equation}
\label{eq:auxiliary}
s = \text{round}(m/C),
\end{equation}
which is used as input for IHT and CoSaMP. Moreover, we consider a two-level structure defined by
$$
\bm{M} = (s/2, N), \quad \bm{s} =(s/2, s/2),
$$
which are employed as input parameters for IHTL and CoSaMPL. To numerically solve BP, we utilize the function \texttt{spg\_bp} from the Matlab toolbox SPGL1 \cite{BergFriedlander:2008, spgl1:2007} with parameters $\texttt{bpTol} = 10^{-6}$, $\texttt{optTol} = 10^{-6}$ and $\texttt{iterations}=1e6$. The IHT(L) and CoSaMP(L) are run with tolerance on the relative increment equal to $10^{-8}$ and a maximum number of 1000 iterations. Moreover, we use a $\sqrt{m/N}$ rescaling of $A$  for IHT(L), as in Section~\ref{sec:Experiments} and let $x^{(0)} = 0$.

\begin{figure}[ht]
\centering
\includegraphics[width = 7cm]{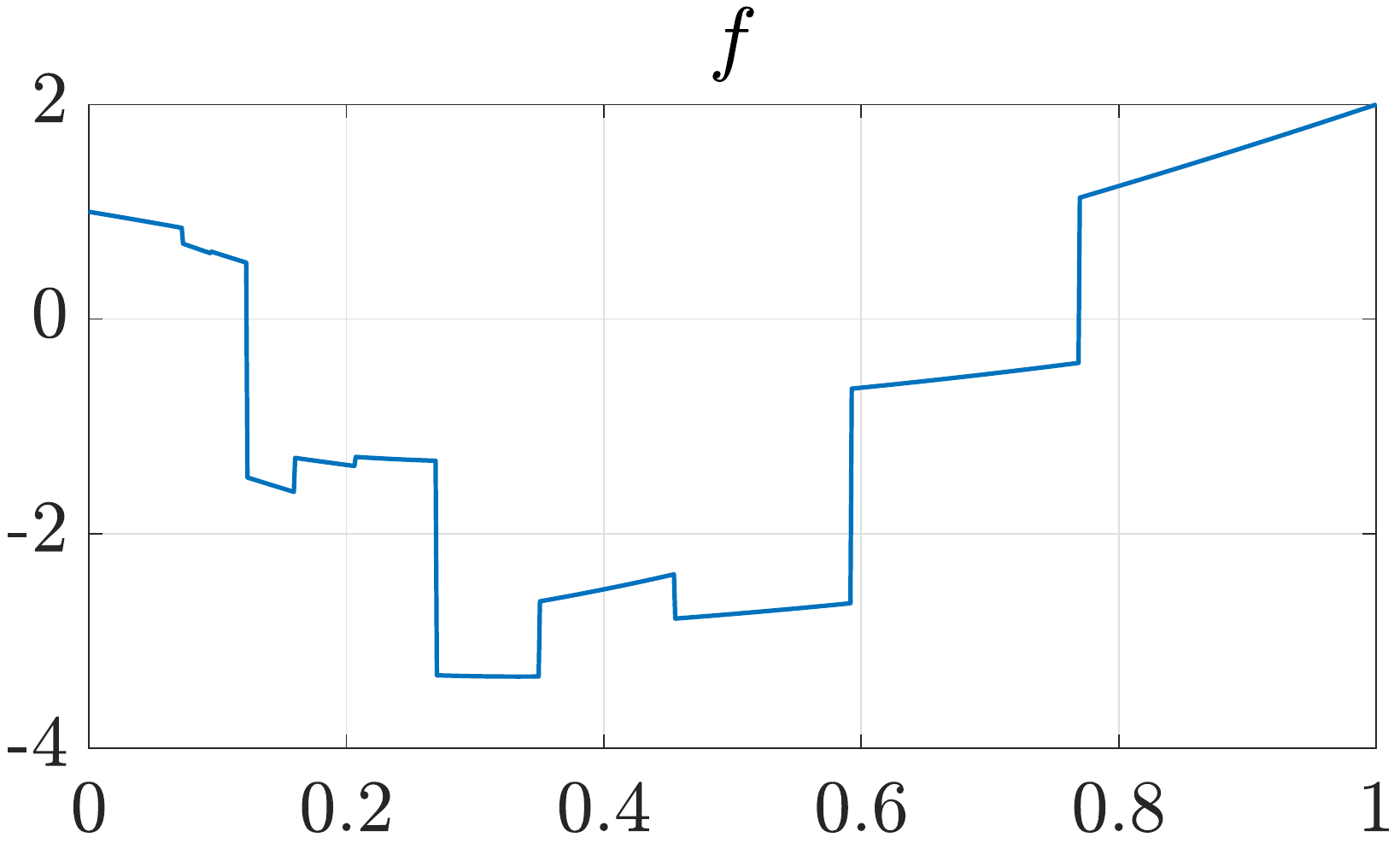}
\caption{\label{fig:f}Piecewise smooth function $f$ defined in \eqref{eq:def_f}.}
\end{figure}

We consider a piecewise smooth function with 10 discontinuities:
\begin{equation}
\label{eq:def_f}
f(x) = 
\sum_{i = 1}^{10} (-1)^{\text{mod}(i, 5)} \; x^{\text{mod}(i,3)} \; \text{sign}(x-(1.3)^{i-9}), \quad 0 \leq x \leq 1.
\end{equation}
Its plot is shown in Figure~\ref{fig:f}. 

We compare all the encoder-decoder pairs for $N = 2^{13}$ and for $m = 2^4, 2^5, \ldots, 2^9$. We plot the relative $L^2$ error as a function of $m$ in Figures~\ref{fig:m_vs_err_IHT(L)} and \ref{fig:m_vs_err_CoSaMP(L)} for values of the auxiliary parameter $C = 3,4,5,10$. The results are averaged over 25 runs. 
\newcommand{\figureheight}{4cm}
\begin{figure}[ht]
\centering
\begin{tabular}{cccc}
Gaussian, IHT \emph{vs.}\ BP & Fourier, IHT \emph{vs.}\ BP \\
\includegraphics[height = \figureheight]{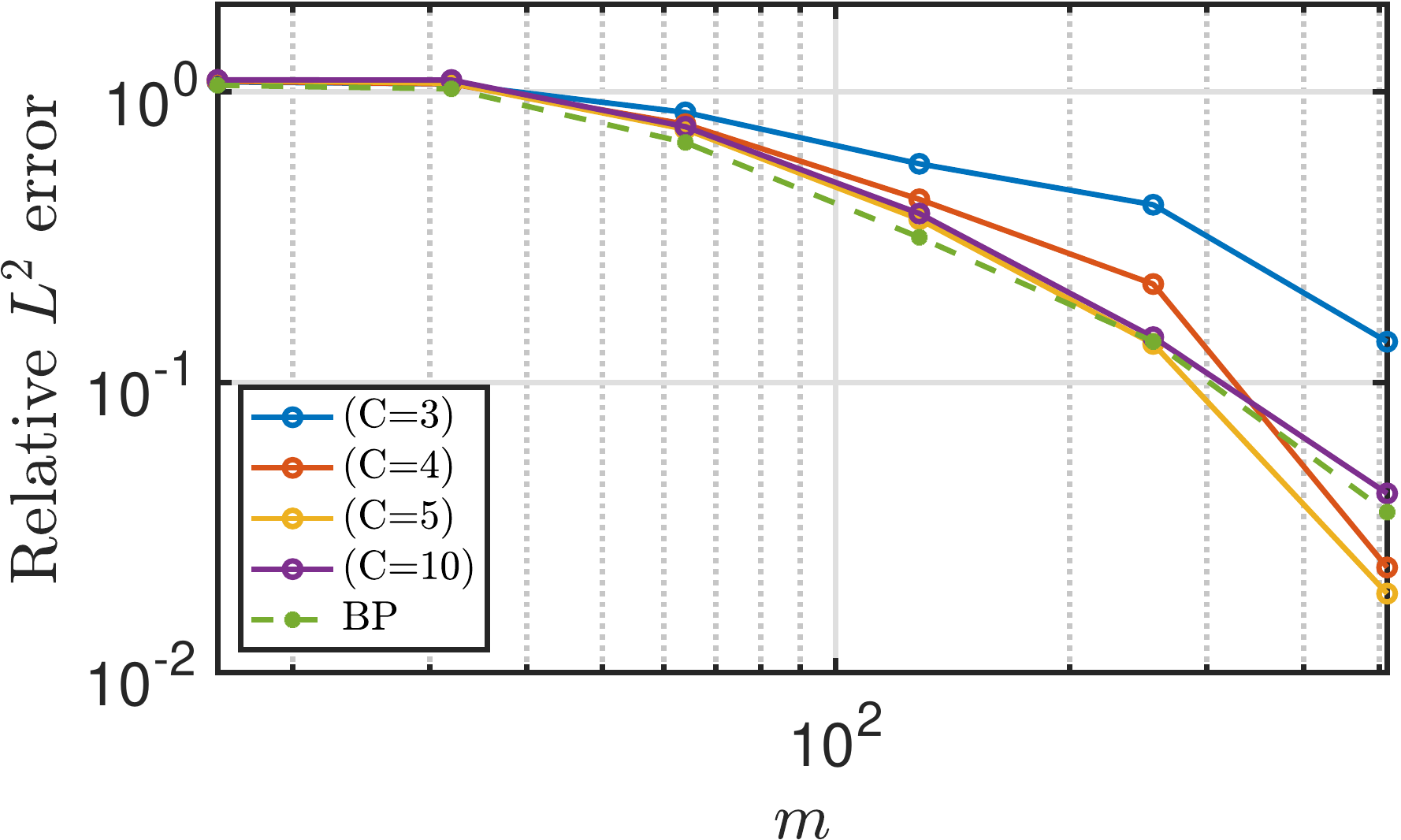} & 
\includegraphics[height = \figureheight]{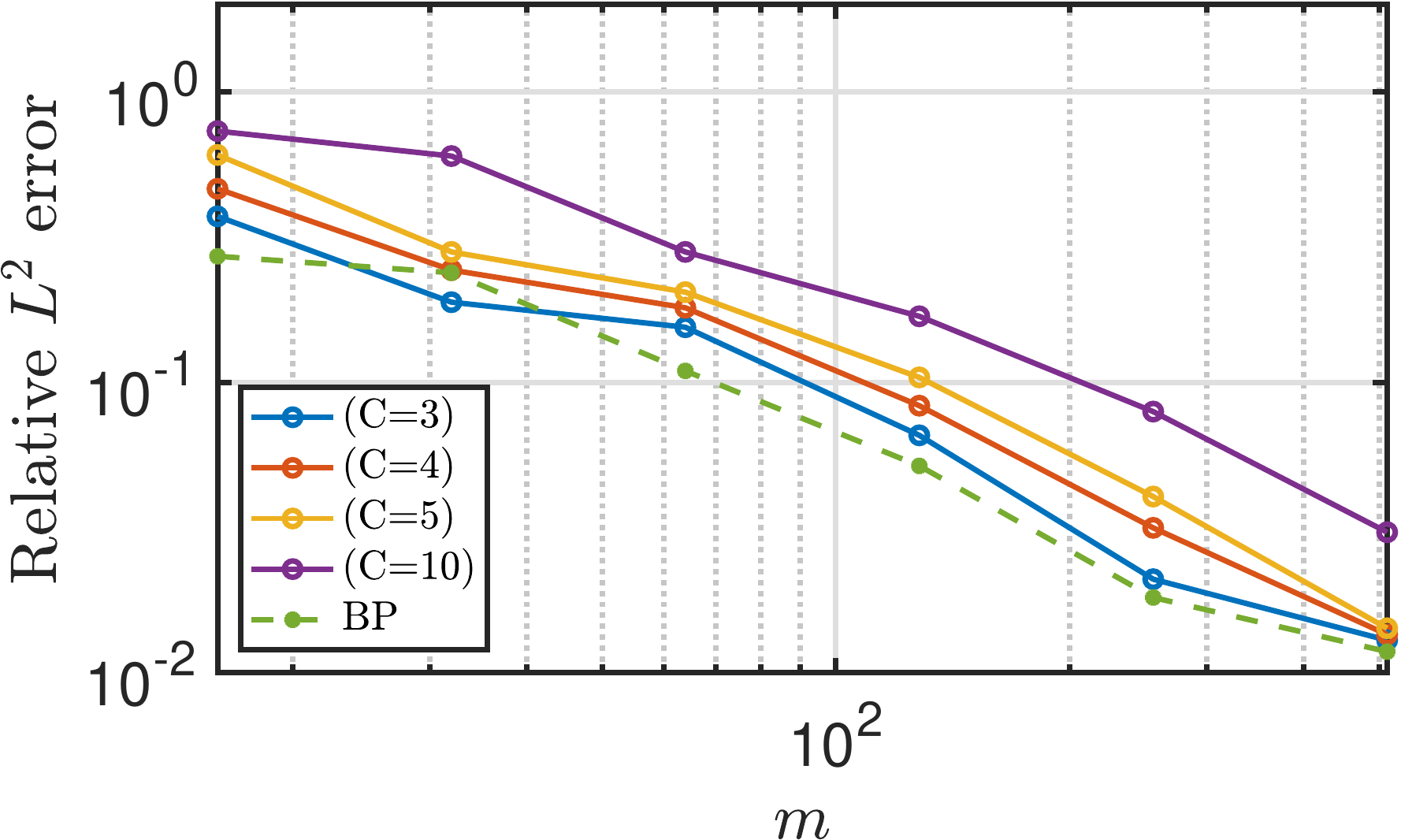}\\
Gaussian, IHTL \emph{vs.}\ BP  & Fourier, IHTL \emph{vs.}\ BP  \\
\includegraphics[height = \figureheight]{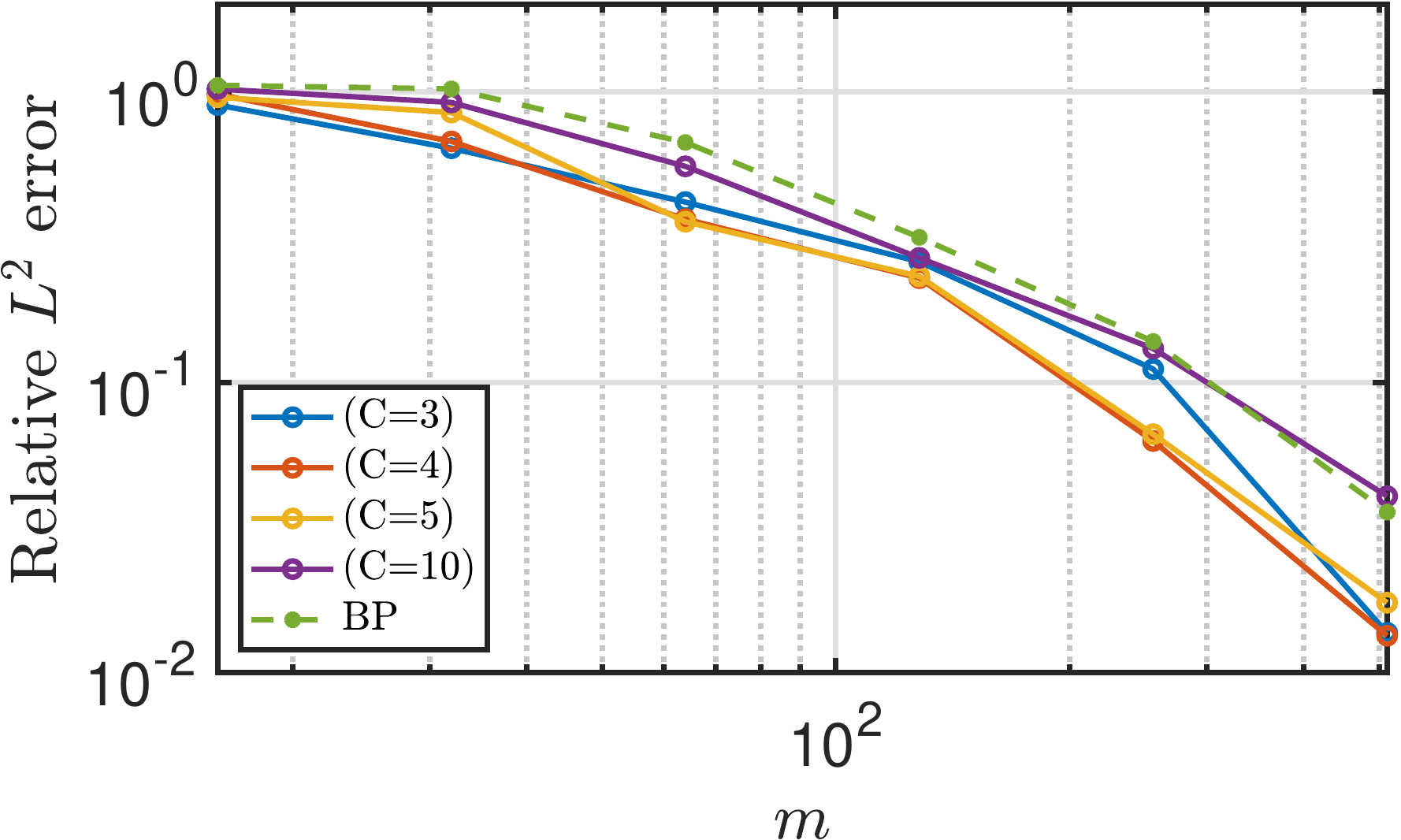} &
\includegraphics[height = \figureheight]{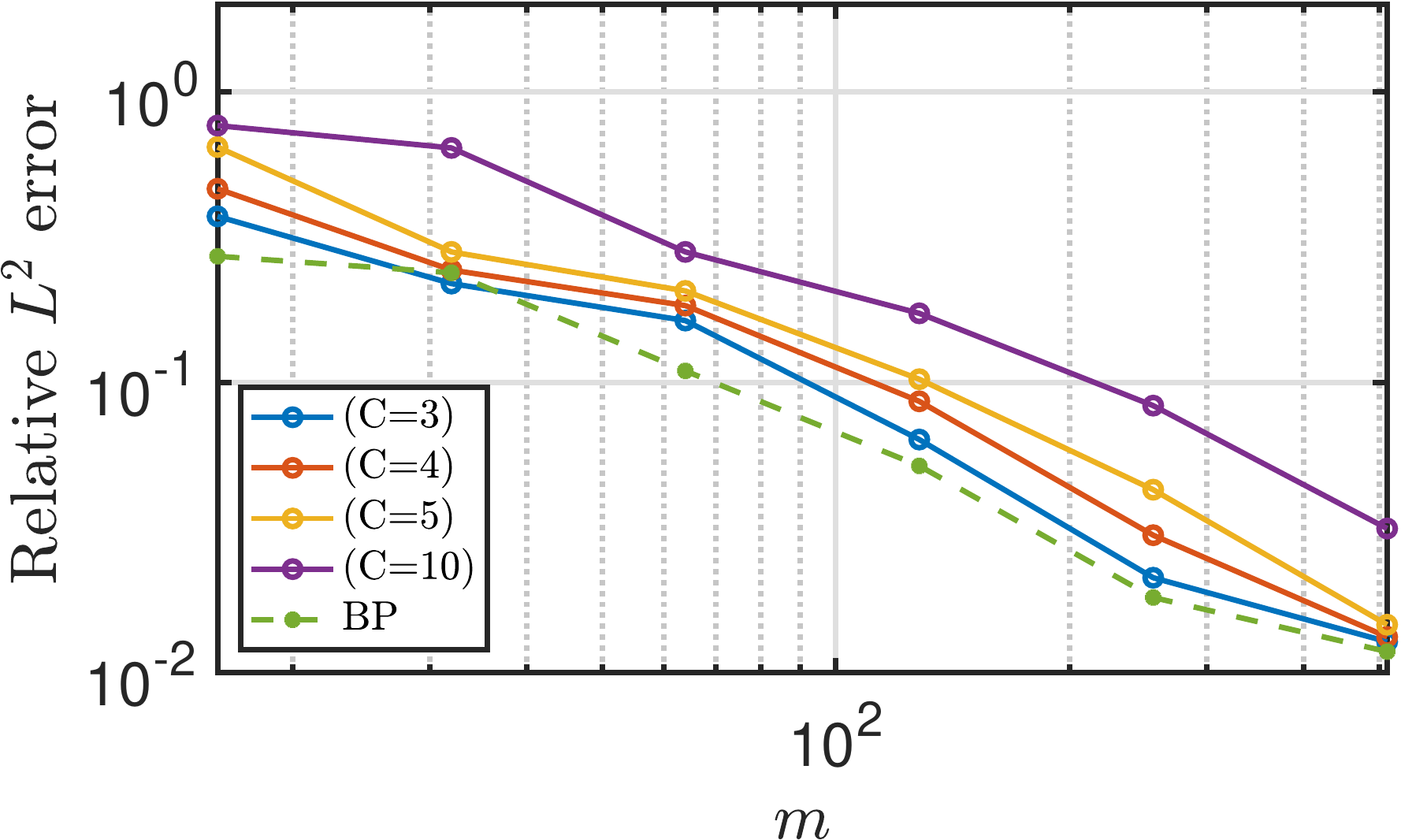}
\end{tabular}
\caption{\label{fig:m_vs_err_IHT(L)} Plot of the relative $L^2$ error vs.\ number of measurements $m$ for the approximation of \eqref{eq:def_f} via different encoder-decoder pairs. Left: Structure agnostic Gaussian encoder. Right: Structure promoting Fourier encoder are consider. Top: The IHT decoder is compared with BP. Bottom: The IHTL decoder is compared with BP.}
\end{figure}

\begin{figure}[ht]
\centering
\begin{tabular}{cccc}
Gaussian, CoSaMP \emph{vs.}\ BP & Fourier, CoSaMP \emph{vs.}\ BP \\
\includegraphics[height = \figureheight]{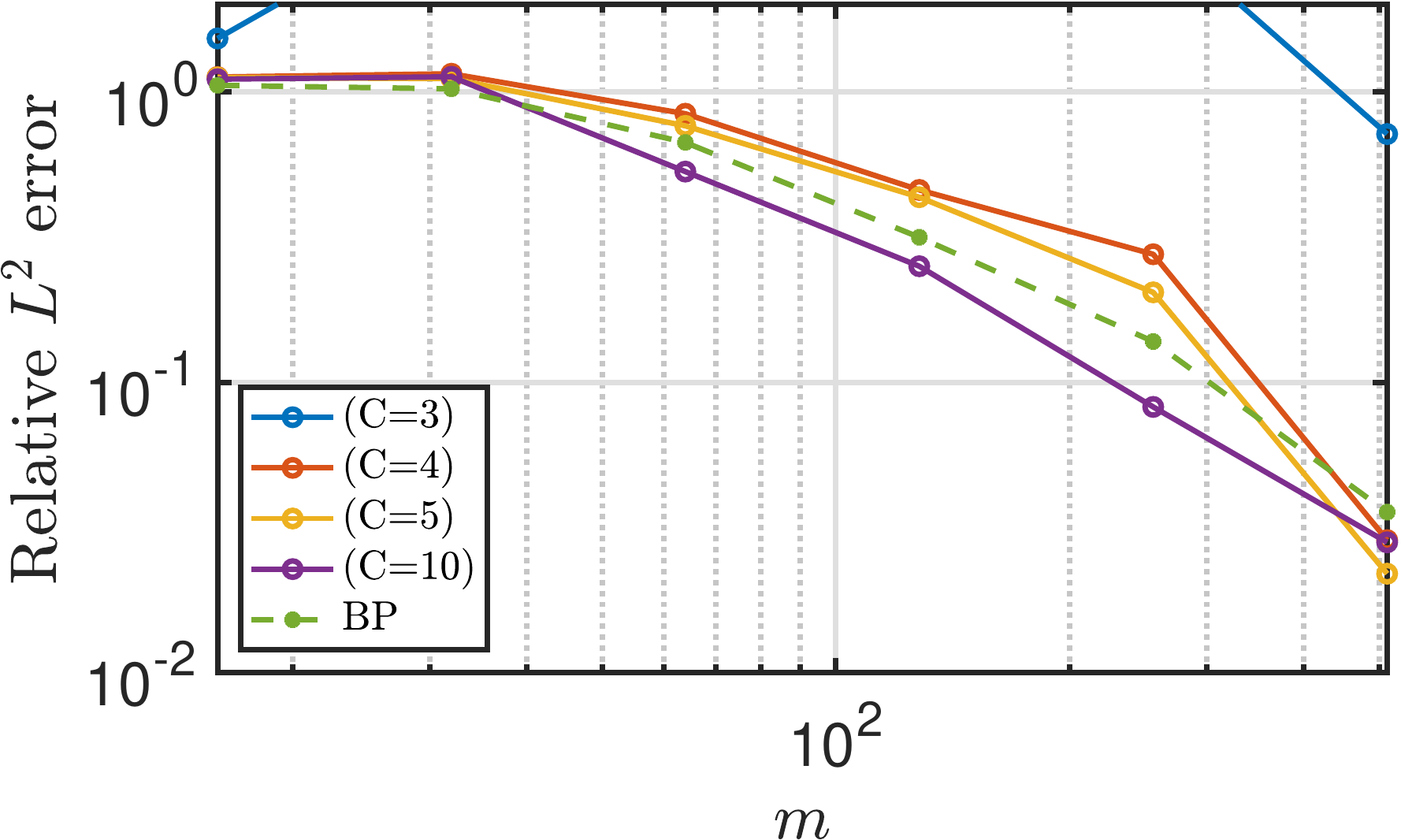} & 
\includegraphics[height = \figureheight]{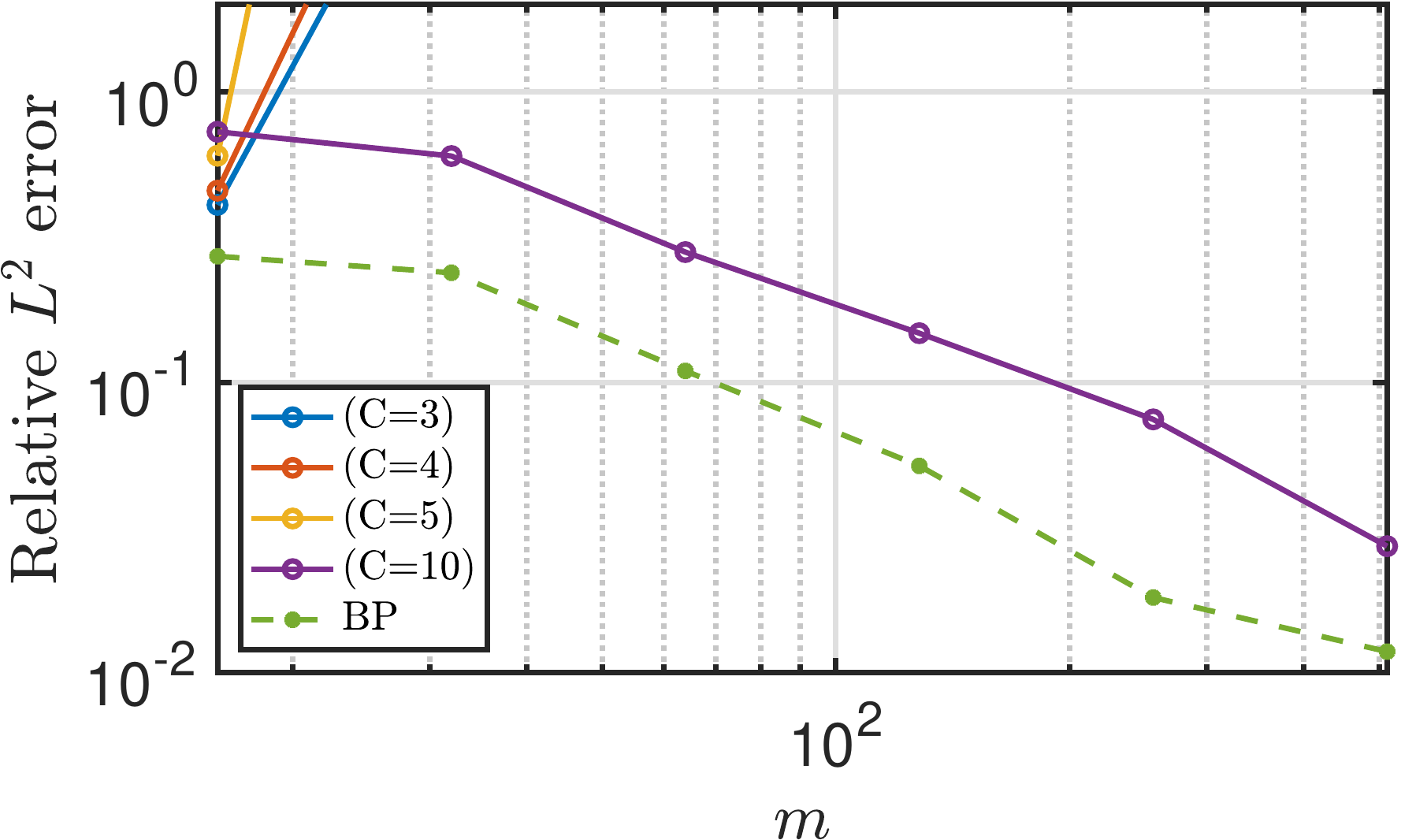} \\
 Gaussian, CoSaMPL \emph{vs.}\ BP &  Fourier, CoSaMPL \emph{vs.}\ BP \\
\includegraphics[height = \figureheight]{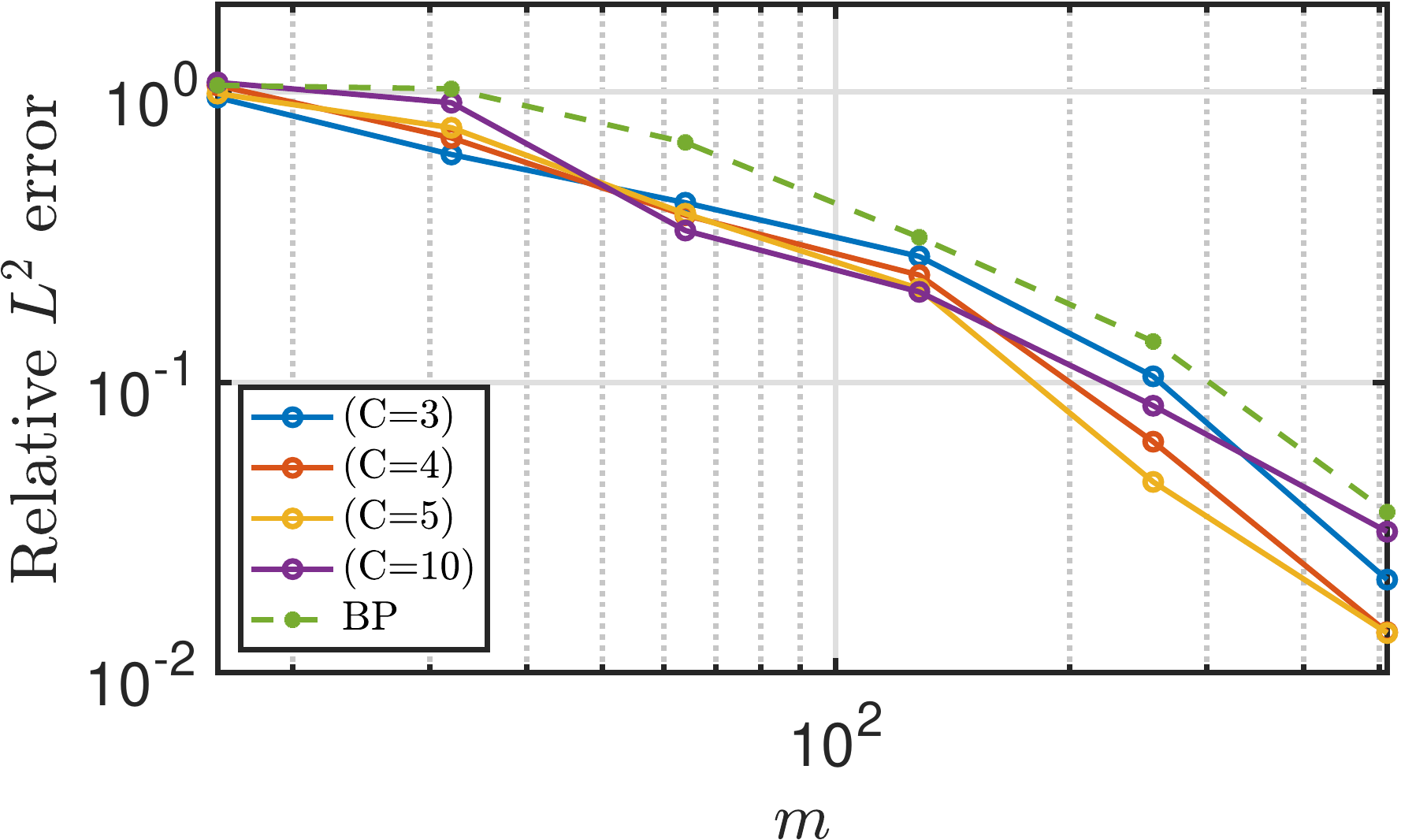} &
\includegraphics[height = \figureheight]{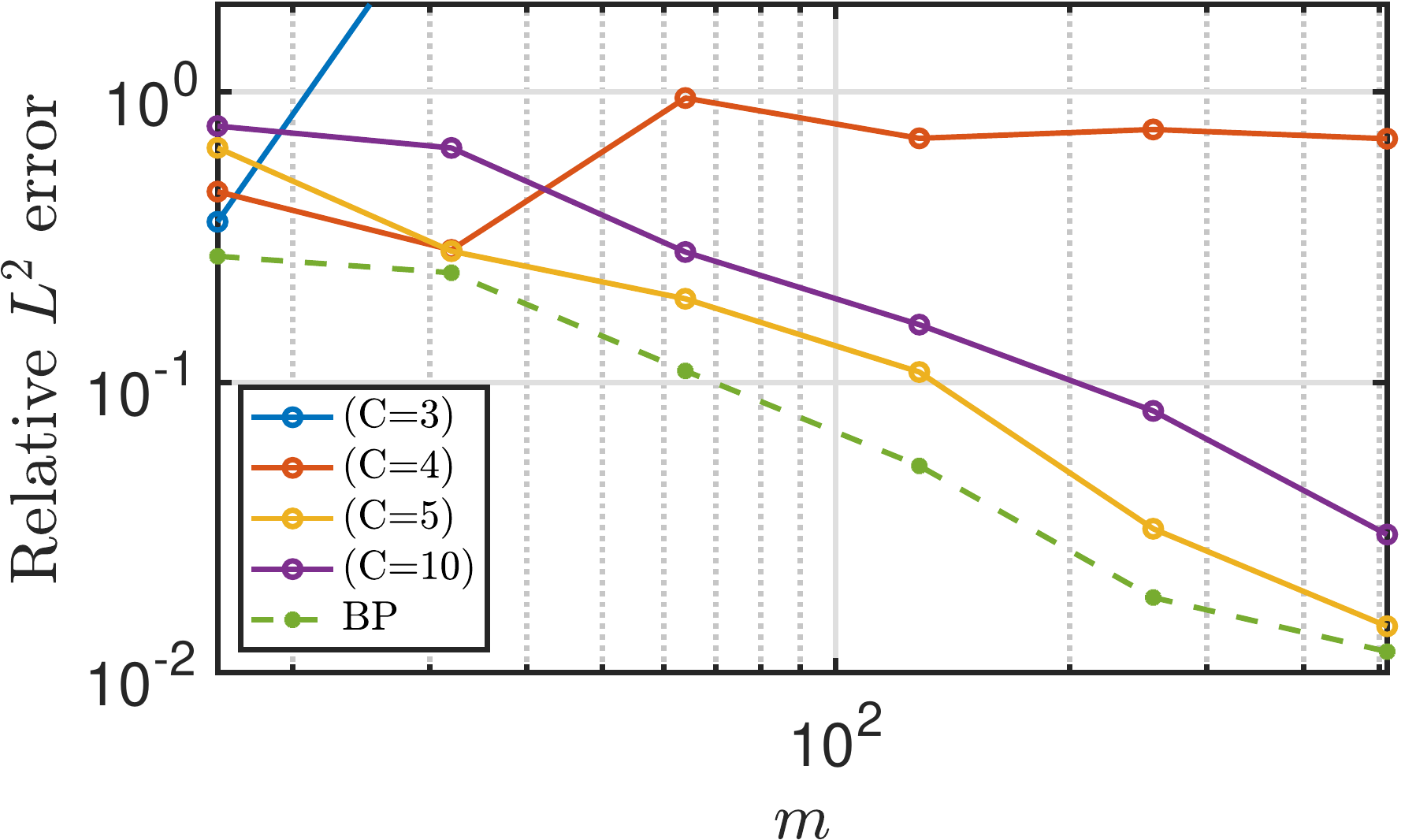} 
\end{tabular}
\caption{\label{fig:m_vs_err_CoSaMP(L)} The same experiment as in Figure~\ref{fig:m_vs_err_IHT(L)} where IHT and IHTL are repaced with CoSaMP and CoSaMPL.}
\end{figure}

In all the experiments, adding the structure in levels to IHT or CoSaMP leads to improved or, in the worst case, comparable approximation accuracy. In particular, in the case of the structure-promoting Fourier encoder, neither IHTL nor CoSaMPL are able to outperform BP or IHT, but we observe that CoSaMPL is more robust than CoSaMP with respect to the choice of $C$. This leads to an interesting conclusion: enforcing structure via the encoder \textit{and} the decoder at the same time is seemingly redundant and does not lead to any additional benefit. We also note that CoSaMP and CoSaMPL are more sensitive to variations of the auxiliary parameter $C$ than IHT and IHTL. On the other hand, in the case of the structure-agnostic Gaussian encoder, we consistently witness the benefits of promoting the sparsity in levels structure in the decoder. Indeed, IHTL and CoSaMPL consistently outperform their unstructured variants and BP.


\section{Conclusions and future work}

We proposed two variants of the IHT and CoSaMP algorithms that promote sparse in levels signals, respectively called IHTL and CoSaMPL. Our numerical experiments show that IHTL and CoSaMPL outperform their unstructured variants when the unknown signal is sparse in levels and, especially, in the case where local sparsities are not uniformly distributed among the levels. The benefits of using a sparsity-in-levels decoder have also been shown in the case of function approximation via compressed sensing, which originally motivated this work. When a structure-promoting encoder based on multilevel Fourier sampling is employed, sparse-in-levels decoders are only able to achieve the same accuracy as $\ell^1$ minimization, but not to outperform it. However, the CoSaMPL and IHTL decoders are able to outperform CoSaMP, IHT, and $\ell^1$ minimization when a structure-agnostic encoder based on random Gaussian sampling is employed.  

The theoretical analysis of stable and robust recovery guarantees for IHTL and CoSaMPL will be presented in a subsequent paper. From the numerical viewpoint, open problems include, e.g.,  the study of adaptive strategies to update the step size in IHTL\cite{BlumensathDavies2010,Blumensath2012} and devising recipes for the automatic choice of the auxiliary parameter $C$ used in \eqref{eq:auxiliary}.  Moreover, a further topic of investigation is the generalization of other greedy and iterative methods, such as the orthogonal matching pursuit algorithm, to the sparsity in levels case.

\acknowledgments 
 
The authors extend their thanks to Kateryna Melnykova for useful suggestions and comments.  S.B.\ acknowledges the support of the PIMS Postdoctoral Training Centre in Stochastics. This work was supported by the PIMS CRG in ``High-dimensional Data Analysis'' and by NSERC through grant R611675.

\bibliography{report} 
\bibliographystyle{spiebib} 

\end{document}